\documentclass[twocolumn,aps,amsfonts,amsmath,prd,nofootinbib]
{revtex4-1}
\usepackage[T1]{fontenc}
 \usepackage[latin1]{inputenc}
\usepackage{amssymb}
\usepackage{amsmath}
\usepackage{enumerate}
\usepackage{epsfig}
\usepackage[pdftex]{hyperref}

\def\K{K{\"a}hler}
\newcommand{\be}{\begin{eqnarray}}
\newcommand{\ee}{\end{eqnarray}}
\def\ba{\begin{array}}
\def\ea{\end{array}}

\makeatletter

\def\ba{\begin{eqnarray}}
\def\ea{\end{eqnarray}}
\def\beas{\begin{eqnarray*}}
\def\eeas{\end{eqnarray*}}
\def\nn{\nonumber}
\def\sla{\raise.15ex\hbox{$/$}\kern-.57em}

\newcommand{\Mpl}{M_{p}}
\newcommand{\edb}{\vec E \cdot \vec B}
\newcommand{\ex}[1]{\langle #1 \rangle}
\newcommand{\intd}[1]{\int \frac{d^{3} #1}{(2\pi)^{3}} }

\begin{document}

\title{ \Large\bf Gauge field production in SUGRA inflation: \\ local non-Gaussianity and primordial black holes}

\author{Andrei Linde$^{1}$}
\author{Sander Mooij$^{2}$}
\author{Enrico Pajer$^{3}$}

\affiliation{$^{1}$Department of Physics, Stanford University, Stanford, CA 94305, USA\\
$^{2}$Nikhef, Science Park 105, 1098 XG Amsterdam, The Netherlands\\
$^{3}$ Department of Physics, Princeton University, Princeton, NJ 08544, USA}


\begin{abstract}
When inflation is driven by a pseudo-scalar field $\chi$ coupled to vectors as ${\alpha\over 4}\,\chi F \tilde F$, this coupling may lead to a copious production of gauge quanta, which in turns induces non-Gaussian and non-scale invariant corrections to curvature perturbations. We point out that this mechanism is generically at work in a broad class of inflationary models in supergravity hence providing them with a rich set of observational predictions.
When the gauge fields are massless, significant effects on CMB scales emerge only for relatively large $\alpha$. We show that in this regime, the curvature perturbations produced at the last stages of inflation have a relatively large amplitude that is of the order of the upper bound set by the possible production of primordial black holes by non-Gaussian perturbations. On the other hand, within the supergravity framework described in our paper, the gauge fields can often acquire a mass through a coupling to additional light scalar fields. Perturbations of these fields modulate the duration of inflation, which serves as a source for non-Gaussian perturbations of the metric. In this regime, the bounds from primordial black holes are parametrically satisfied and non-Gaussianity of the local type can be generated at the observationally interesting level $f_{\rm NL}\sim \mathcal{O}(10)$. 
\end{abstract}

\begin{preprint}
 .NIKHEF 2012-024
\end{preprint}

\maketitle

\section{Introduction} \label{intro}

In a recent series of papers \cite{Kallosh:2010ug,Kallosh:2010xz,Kallosh:2011qk} a new broad class of models of chaotic inflation in supergravity has been developed. These models generalize the simplest model of this type proposed  long ago in \cite{Kawasaki:2000yn}; see also \cite{Yamaguchi:2000vm,Yamaguchi:2001pw,Kawasaki:2001as,Yamaguchi:2003fp,Brax:2005jv,Kallosh:2007ig,Kadota:2007nc,Davis:2008fv,Einhorn:2009bh,Ferrara:2010yw,Lee:2010hj,Ferrara:2010in,Takahashi:2010ky,Nakayama:2010kt,Silverstein:2008sg,McAllister:2008hb}
 for a partial list of other closely related publications. 
 
 The new class of models \cite{Kallosh:2010ug,Kallosh:2010xz,Kallosh:2011qk} describes two scalar fields, $S$ and $\Phi$, with the superpotential 
 \be
{W}= Sf(\Phi) \ ,
\label{cond}
\ee
where $f(\Phi)$ is a real holomorphic function such that $\bar f(\bar \Phi) = f(\Phi)$. Any function which can be represented by Taylor series with real coefficients has this property. The \K \, potential can be  chosen to have functional form
\be\label{Kminus}
K= K((\Phi-\bar\Phi)^2,S\bar S).
\ee
In this case, the \K\, potential does not depend on $\phi = \sqrt 2\, {\rm Re}\, \Phi$. Under certain conditions on the \K\, potential, inflation occurs along the direction $S = {\rm Im}\, \Phi = 0$, and the field $\phi$ plays the role of the inflaton field with the potential 
\be
V(\phi) = |f(\phi/\sqrt 2)|^{2}.
\ee
 All scalar fields have canonical kinetic terms along the inflationary trajectory $S = {\rm Im}\, \Phi = 0$. 
 
This class of models can be further extended \cite{Davis:2008fv,Kallosh:2011qk} to incorporate a KKLT-type construction with strong mo\-du\-li stabilization  \cite{Kallosh:2004yh}, which may have interesting phenomenological consequences and may provide a simple solution of the cosmological moduli and gravitino problems  \cite{Linde:2011ja,Dudas:2012wi}.

The generality of the functional form of the inflationary potential $V(\phi)$ allows one to describe {\it any} combination of the parameters $n_{s}$ and $r$. Thus, this rather simple class of models may describe {\it any} set of observational data which can be expressed in terms of these two parameters by an appropriate choice of the function $f(\Phi)$ in the superpotential. Meanwhile the choice of the \K\, potential controls masses of the fields orthogonal to the in\-fla\-ti\-o\-na\-ry trajectory \cite{Kallosh:2010ug,Kallosh:2010xz,Kallosh:2011qk}. Reheating in this scenario requires considering the scalar-vector coupling $\sim  \phi F_{\mu \nu}F^{\mu\nu}$ \cite{Kallosh:2011qk,Ferrara:2011dz}.
If not only the inflaton but some other scalar field has a mass much smaller than $H$ during inflation, one may use it as a curvaton field   \cite{curva} for generation of non-Gaussian perturbations in this class of models \cite{Demozzi:2010aj}. 

In this paper, we will study an alternative formulation of this class of models, with 
the \K\, potential \be\label{Kplus}
K= K((\Phi+\bar\Phi)^2,S\bar S).
\ee
The simplest version of models of that type, 
with ${W}= m S\Phi$ the \K\, potential $
\mathcal{K} = S \bar S + \frac{1}{2}(\Phi+\bar\Phi)^2$
was first proposed \cite{Kawasaki:2000yn}. In this class of models, the \K\, potential does not depend on $\chi = \sqrt 2\, {\rm Im }\, \Phi$, which plays the role of the inflaton field with the potential 
\be
V(\chi) = |f(\chi/\sqrt 2)|^{2}.
\ee
The description of inflation in the models (\ref{Kminus}) and (\ref{Kplus}) coincides with each other, up to a trivial replacement $\phi \to \chi$, as long as vector fields are not involved in the process.

The difference appears when one notices that in the model (\ref{Kplus}) the inflaton field is a pseudo scalar, which can have a coupling to vector fields
\be
{\alpha\over 4}\, \chi F_{\mu \nu}\tilde{F}^{\mu\nu},
 \ee
where $\tilde{F}^{\mu\nu}\equiv \epsilon^{\mu\nu \rho\sigma}F_{\rho\sigma}$ and $\alpha$ is a dimensionful constant. This coupling is expected to be present since it is compatible with all the symmetries, including a shift-symmetry in $\chi$.

 The study of the phenomenological effects of such a coupling during inflation has received a lot of attention lately \cite{Anber:2006xt,Anber:2009ua,Durrer:2010mq,Barnaby:2010vf,Barnaby:2011vw,Barnaby:2011qe,Meerburg:2012id}. In particular, it has been shown in  \cite{Barnaby:2010vf,Barnaby:2011vw} that, if the constant $\alpha$ is large enough, such a coupling can lead to a copious production of gauge fields due to the time dependence of $\chi$. Through their \textit{inverse decay} into inflaton perturbations, these gauge fields yield an additional contribution to the scalar power spectrum which is both non-Gaussian and violates scale invariance. In this way it is possible to obtain non-Gaussian and non-scale invariant effects that can be observed by the Planck satellite and has not yet been ruled out yet by WMAP, although the parameter space corresponding to such a signal is relatively small \cite{Meerburg:2012id}. In addition, gauge fields source tensor modes and lead to a stochastic gravity wave signal that could be detected at interferometers such as Advanced LIGO or Virgo \cite{SC,Barnaby:2011qe} (see also \cite{AS}).
 
Since the new class of inflationary models in supergravity needs a coupling between the inflaton and gauge fields to have successful reheating, we have to consistently take into account the violations of Gaussianity and scale invariance induced by the inverse decay mechanism. This is the topic of section \ref{nongausssection}.

A potential threat in this model is the overproduction of primordial black holes. As we will see in section \ref{bk}, at very small scales, far beyond what is observable by the CMB, the produced gauge quanta largely increase the curvature power spectrum. At some point, various forms of backreaction stops this growth, but by then the power spectrum has reached $\Delta_\zeta^{2}\sim \mathcal{O}(10^{-3})$. At such high values, a statistical fluctuation might locally increase the density so that primordial black holes are formed. In this way the non-detection of primordial black holes puts an observational upper bound on the power spectrum \cite{malik,Carr:2009jm,Byrnes:2012yx,Klimai:2012sf,Lyth:2012yp,Shandera:2012ke}, which we discuss in section \ref{BH}. Our estimates for the late-time power spectrum land a factor of six above this bound (compare e.g.~\eqref{finalps} with \eqref{finbound}). Since we expect our estimate to be reliable up to factors of order one, we cannot definitively claim that the inverse decay mechanism and its interesting phenomenology is incompatible with current data, but our result on production of primordial black holes highlights a clear tension.

In section \ref{mA} we describe an alternative mechanism of generation of non-Gaussian perturbations, proposed in \cite{Meerburg:2012id}. This mechanism requires introduction of a light charged field $h$ with mass $m_{h} \ll H$, where $H$ is the Hubble constant during inflation.  Inflationary perturbations of this field generate a slightly inhomogeneous distribution of a classical scalar field $h(x)$. This field induces the vector field mass due to Higgs effect.

As a result, the vector field mass $\sim eh(x)$ takes different values, controlled by fluctuations of the field $h$.   In the parts of the universe where the value of the vector field mass is small, the vector field fluctuations are easily produced since the gauge mass quenches the tachyonic instability. This in turns leads to a longer stage of inflation because of the additional friction generated by the gauge fields.  Meanwhile in the parts of the universe where the fluctuations of the light scalar field $h$ make this field large, the vector field mass becomes larger and inflation is shorter due to the lack of backreaction. As a result, fluctuations of the light scalar field $h$ lead to fluctuations of the total number of e-foldings $\delta N$, i.e. to adiabatic perturbations of metric. We will show that this effect may generate significant primordial local non-Gaussianity.  Also, in the regime of parameters relevant for this scenario the primordial black hole bounds are satisfied parametrically. 

To implement this mechanism in our supergravity-based inflationary scenario, one should find a way to guarantee smallness of the mass of the field $h$ during inflation.   We will describe a model where the mass squared of this field during inflation is equal to $m^{2}_{h}=\gamma H^{2}$, where $\gamma$ can be made small by a proper choice of the \K\, potential.

In section \ref{stoch} we study the evolution of the light field $h$ during inflation in our scenario, which is similar to the evolution of the curvaton field $\sigma$ in \cite{Demozzi:2010aj}, so for simplicity we will continue calling this field the curvaton. One can use the results of \cite{Demozzi:2010aj} for the description of its evolution. However, in the original model of  \cite{Demozzi:2010aj}, just as in any other curvaton model   \cite{curva}, adiabatic perturbations of metric are generated by  perturbations of the field $\sigma$ after a complicated sequence of reheating, expansion of the universe, and the subsequent decay of the curvaton field. In our scenario, adiabatic perturbations are produced due to the modulation of the duration of inflation by the perturbations of the field $h$.
As we will demonstrate, this mechanism can easily produce local non-Gaussianity in the potentially interesting range $f_{{\rm NL}}$ from  $\mathcal{O}(10)$ to $\mathcal{O}(100)$, even if the coupling constant $\alpha$ is not very large. 

Finally, in section \ref{reh}, we find that typical values of the coupling constant $\alpha$ considered in this work lead to a relatively high perturbative reheating temperature $T \sim 10^{10} {\rm GeV}$. This should be read as a lower limit, since the copious non-perturbative production of gauge fields already during inflation could lead to and even higher reheating temperature. This could lead to the cosmological gravitino problem \cite{gravitino}, but in the class of models with strong moduli stabilization and gravitino mass $\mathcal{O}(100)$ TeV this problem does not appear  \cite{Dudas:2012wi}.


\section{CMB scales: violations of Gaussianity and scale invariance} \label{nongausssection}

Recently there has been a lot of interest in the effect of gauge field production in axion inflation \cite{Anber:2006xt,Anber:2009ua,Durrer:2010mq,Barnaby:2010vf,Barnaby:2011vw,Barnaby:2011qe,Meerburg:2012id}. In this section we summarize the main points. 

Consider a pseudo-scalar inflaton with a potential suitable for inflation. The symmetries of the theory allow for a coupling $ \chi F_{\mu \nu}\tilde F^{\mu\nu}$  to some $U(1)$ gauge sector. This coupling is essential for reheating in the supergravity models we discussed in section \ref{intro}. We will therefore consider the following bosonic part of the action\footnote{Notice that in the existing literature, such a coupling is usually associated with interaction of the axion field with vector fields, with a coupling $-{\alpha\over 4 f}$. In our approach it is not necessary to associate the pseudo scalar field with the axion field with the radius of the potential $\sim f$, so we normalize the coupling in terms of the reduced Planck mass $\Mpl$, which we then set to one, and consider the following interaction term $-{\alpha \over 4} \chi F_{\mu \nu}\tilde F^{\mu\nu}$.}
\be
S=-\int d^{4}x \sqrt{-g}\left[\frac12 \left(\partial \chi\right)^{2}+\frac14 F^{2}+\frac{\alpha}{4} \chi F\tilde F+V(\chi)\right]\,.\nonumber
\ee
Since all relevant effects arise from the couplings above we can safely neglect the gravitational interaction between perturbations and work with an unperturbed FLRW metric\footnote{We are neglecting vector and tensor modes and the slow-roll suppressed interactions coming from the solution of the constraints on the lapse and the shift.}. We organize the perturbation theory based on the equations that we are able to solve. Consider two classical\footnote{Here we are assuming that the occupation number of the relevant gauge modes is large enough that one can approximate the resulting electromagnetic field with a classical one. This assumption is implicit in all other approaches so far.} fields $\vec{A}(x,t)$ and $\chi(t)$ that solve these two coupled differential equations
\be
\ddot \chi+3H\dot \chi+\frac{\partial V}{\partial\chi}&=&\alpha \ex{\edb}\,, \label{chieq}\\
\vec{A}''-\nabla^{2}\vec{A}-\alpha\chi'\,\nabla \times \vec{A}&=&0\,,\label{ab}
\ee
where $\vec E \equiv -  \dot{\vec A}/a$, $\vec B\equiv a^{-2}\nabla \times \vec A$ and $\edb=-F\tilde F/4$ are computed from $\vec{A}$. 

Now let us look at the action expanded around $\chi$ and $\vec{A}$, i.e. $S[\chi+\delta \chi,\vec A+\delta \vec{A}]$. Organizing the result at various orders in $\delta \chi$ and $\delta \vec{A}$ one finds 
\be
S&=&{\rm const} -\int d^{4}x \sqrt{-g}(\delta \chi)\alpha \left[\ex{\edb}-\edb\right]\,\\
&&-\int d^{4}x \sqrt{-g}\Bigl[\frac12 (\partial \delta \chi)^{2}+\frac12 \frac{\partial^2 V}{\partial \chi^2} (\delta \chi)^{2}+\frac14 (\delta F)^{2}\nonumber\\
&& \qquad \qquad+\frac{\alpha}{4} \chi  \delta F\delta\tilde F+\frac{\alpha}{2} \delta \chi \delta F\tilde F\Bigr]\,\nonumber \\ 
&&-\int d^{4}x \sqrt{-g} \left[\frac{ \alpha}{4} \delta\chi \delta F \delta\tilde{F} + \frac16 (\delta\chi)^{3} \frac{\partial^3 V}{\partial \chi^3} \right]\,,\nonumber 
\ee
where again the classical background fields $\chi$ and $\vec{A}$ solve \eqref{chieq} and \eqref{ab}. Notice that there is a ``tadpole'' for $\delta \chi$ due to the fact that at the background level we solved an inhomogeneous equation for $\vec{A}$ but just a homogeneous one for $\chi$. From this term one also sees that $\delta\chi$ will source $\delta A^{0}$, hence it will modify the constraint. The equations of motion in Coulomb gauge $\partial_{i} A^{i}=0$ are 
\be\label{eom1}
&&a\ddot{\delta A_{i}}-\frac{\partial_{k}^{2}(\delta A_{i})}{a}+aH\dot{\delta A_{i}}-\alpha\dot \chi\nabla\times(\delta\vec{A})=\\
 &&\qquad \alpha \dot{\delta \chi}\nabla\times\vec{A}-\alpha (\nabla \delta\chi)\times \dot{\vec{A}}-\nonumber\partial_{t}(a\partial_{i}(\delta A^{0}))\,,\\ \label{eom2}
&&\ddot{(\delta\chi)}+3H \dot{\delta\chi}-\nabla^{2}\delta \chi+\frac{\partial^2 V}{\partial \chi^2} \delta \chi=\\
 &&\qquad \qquad \qquad\qquad\frac{\alpha}{4}\left(\ex{F\nn\tilde F}-F\tilde F-2\delta F\tilde F\right)\,,\\ \label{eom3}
&& a\partial_{i}\partial_{i}(\delta A)^{0}=-\alpha \nabla (\delta\chi) \cdot \nabla\times\vec{A}\,.
\ee
The solution for the constraint equation for $\delta A^{0}$ is
\be
\delta A^{0}(x,t)=a^{-1}\int d^{3}y \,\frac{\alpha \nabla (\delta\chi) \cdot \nabla\times\vec{A}}{4\pi|x-y|}\,.
\ee 
Unfortunately this coupled system of equations is hard to solve. Hence \cite{Barnaby:2010vf,Barnaby:2011vw} made the approximation of neglecting all terms quadratic or higher in $\delta\chi$, $\delta A$ and $A$, which yields

\begin{eqnarray}
&& \hskip -8pt a\ddot{\delta A}_{i}-\frac{\partial_{k}^{2}\delta A_{i}}{a}+aH\dot{\delta A}_{i}-\alpha \dot{\chi}\nabla\times \delta \vec{A}=0 \label{AB}\\
&& \hskip -8pt \ddot{\delta \chi}+3H \dot{\delta\chi}-\nabla^{2}\delta \chi+\frac{\partial^2 V}{\partial \chi^2} \delta \chi =\nn\\
&& \qquad \qquad \qquad \qquad\qquad\alpha\left(\ex{\edb}-\edb\right)\, .\label{hKG}
\end{eqnarray}

 This is a good approximation as long as $F\tilde F$ (or equivalently $\ex{\edb}$) is not too large (a more quantitative condition is given in \eqref{br2}), which is the regime we will discuss in this section. Note from \eqref{ab} and \eqref{AB} that in this approximation there is no way to tell $A$ apart from $\delta A$.  In the next section we will see that, since $\ex{\edb}$ grows with time, towards the end of inflation this description in not valid anymore, and one has to take backreaction into account.

Solving the equation of motion \eqref{ab} one finds a tachyonic enhancement of the gauge fields. For the growing mode of one of the two polarizations of the gauge field we get
\be
A=\frac{1}{\sqrt{2k}}e^{\pi \xi/2}W_{-i\xi,1/2}(2ik\tau). \label{asol}
\ee
Here $W_{\lambda,\mu}(x)$ denotes the Whittaker function, and $\xi$ is defined as\footnote{Note that we have some minus signs different from \cite{Barnaby:2010vf}, but this is a matter of conventions. We will work with a model that has $\dot{\chi}<0$ during inflation and define $\xi$ to be positive. The sign of $\ex{\vec{E} \cdot\vec{B}}$ is always opposite to the sign of $\dot{\chi}$. Therefore the physical effect of the tachyonic enhancement is always that inflation is prolonged. To be precise: when $\dot{\chi}$ is negative, the growing field is actually the opposite polarization, i.e.~$A_-$, which makes that $\ex{\edb}>0$ (see, for example, equation (8) in \cite{Anber:2009ua}).\label{footnote4} }
\be
\xi \equiv -{\dot\chi\alpha\over 2 H}. \label{xi}
\ee
As we see, the relation between the coupling constant $\alpha$ and the value of $\xi$ 60 e-foldings before the end of inflation is model dependent, but for our model there is an approximate relation which is valid for the parameters that we are going to explore:
\be
\alpha \sim 15 \xi. \label{xi2}
\ee
For $\xi>1$ the new coupling therefore leads to generation of perturbations of the vector fields around horizon scales. The produced gauge fields then change the dynamics of $\chi$ and $H$. The cosmological homogeneous Klein-Gordon equation and the Friedmann equation get extra contributions from the gauge fields and can now be written as
\ba
&&\ddot{\chi}+3H\dot{\chi} +\frac{\partial V}{\partial\chi}= \alpha \ex{\vec{E} \cdot \vec{B}}\label{kg}\\
&&3H^2=\frac{1}{2}\dot{\chi}^2+V+ \frac{1}{2} \ex{\vec{E}^2+\vec{B}^2} \label{fr}.
\ea
They are computed as
\ba
\ex{\vec{E} \cdot \vec{B}}&=& \frac{1}{4\pi^2 a^4} \int_0^\infty dk k^3 \frac{\partial}{\partial \tau}|A|^2,\label{edbdef} \\
\ex{\frac{\vec{E}^2+\vec{B}^2}{2}}&=& \frac{1}{4\pi^2 a^4} \int_0^\infty dk k^2 \Biggl[  |A'|^2 + k^2 |A|^2  \Biggr]. \nonumber\\
&&
\ea
After renormalization, one can reduce the integration interval to the region $\frac{1}{8\xi}< \frac{k}{aH}<2\xi$, which is where the enhancement in the (derivative of the) gauge field takes place.

From the homogeneous Klein-Gordon equation \eqref{hKG} one reads off that the influence of the produced gauge fields on the homogeneous dynamics of $\chi$ and $H$ can be safely neglected as long as
\be
\frac{\alpha\ex{\vec{E} \cdot \vec{B}}}{3H\dot{\chi}}\ll1,\qquad \frac{\frac{1}{2} \ex{\vec{E}^2+\vec{B}^2}}{3H^2}\ll1. \label{br}
\ee
Of these two conditions the first one is always the most stringent. When it stops to hold, backreaction on the homogeneous evolution becomes important and the evolution of $\chi$ and $H$ will be slowed down, which makes inflation lasts longer. We will see in the next section that backreaction on the inhomogeneous equation for $\delta\chi$ happens even earlier. In this section we focus on the regime in which all of these effects are negligible, which e.g.~for a quadratic potential corresponds roughly to $\xi\lesssim 4$. This is appropriate for the description of CMB scales.

Now we move to the power spectrum. The copiously generated gauge fields may, by inverse decay, produce additional perturbations of the inflaton field $\delta \chi$, proportional to the square of the vector field perturbations. As was shown in \cite{Barnaby:2010vf,Barnaby:2011vw}, this can be described (up to backreaction effects to be described in the next section) by using \eqref{hKG}.
The inclusion of the source term leads to an extra contribution to the power spectrum of the curvature perturbation on uniform density hypersurfaces $\zeta=-\frac{H}{\dot{\chi}} \delta\chi$, which has been computed in \cite{Barnaby:2010vf,Barnaby:2011vw} (we present a quick estimate in Appendix \ref{turbops})
\be\label{spectrum}
\Delta^2_{\zeta}(k) = \Delta^2_{\zeta,{\rm sr}}(k) \left(1 + \Delta^2_{\zeta,{\rm sr}}(k)\, f_{2}(\xi) e^{4\pi\xi}\right),
\ee
where $f_{2}(\xi)$ was defined in \cite{Barnaby:2010vf,Barnaby:2011vw} and can be computed numerically (a useful large $\xi$
approximation is given in \eqref{f2}) and 
\be
\Delta^2_{\zeta,{\rm sr}}(k)  = \left({H^{2}\over 2\pi |\dot\chi|} \right)^2
\ee
is the amplitude of the vacuum inflationary perturbations as in standard slow-roll inflaton.  WMAP \cite{Komatsu:2010fb} has measured $\Delta^2_{\zeta,{\rm sr}}(k_\star)=2.43\cdot 10^{-9} $, where $k_\star=0.002{\rm Mpc}^{-1}$ is the pivot scale that we will take to correspond with $N=60$ e-foldings before the end of inflation. The second term in \eqref{spectrum} violates both scale invariance (and Gaussianity as we will see below), since it comes schematically from $A^{2}$, i.e.~the square of a Gaussian which grows with time as in \eqref{asol}.

\begin{figure}[t!]
\begin{center}
\includegraphics[scale=0.35]{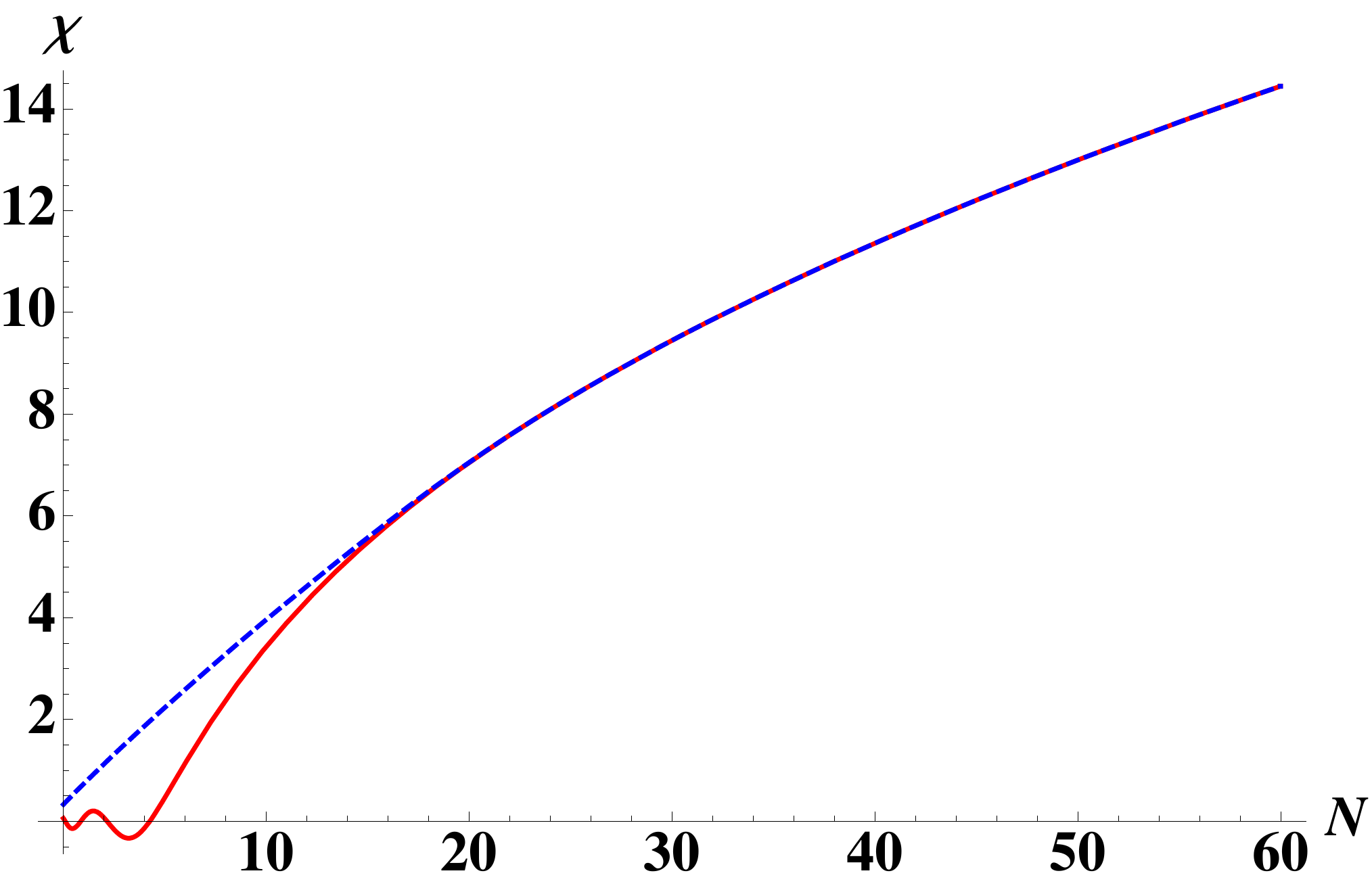}
\end{center}
\caption{\footnotesize The evolution of the inflaton field $\chi$, as a function of the number of e-folds $N$ left to the end of inflation (time is moving to the left) for $\xi[N=60]=2.2$. The result in dashed blue does take backreaction from the sources in equations \eqref{kg} and \eqref{fr} into account, the result in red does not. It is clear that backreaction prolongs inflation.}
\label{chi}
\end{figure}

\begin{figure}[t!]
\begin{center}
\includegraphics[scale=0.35]{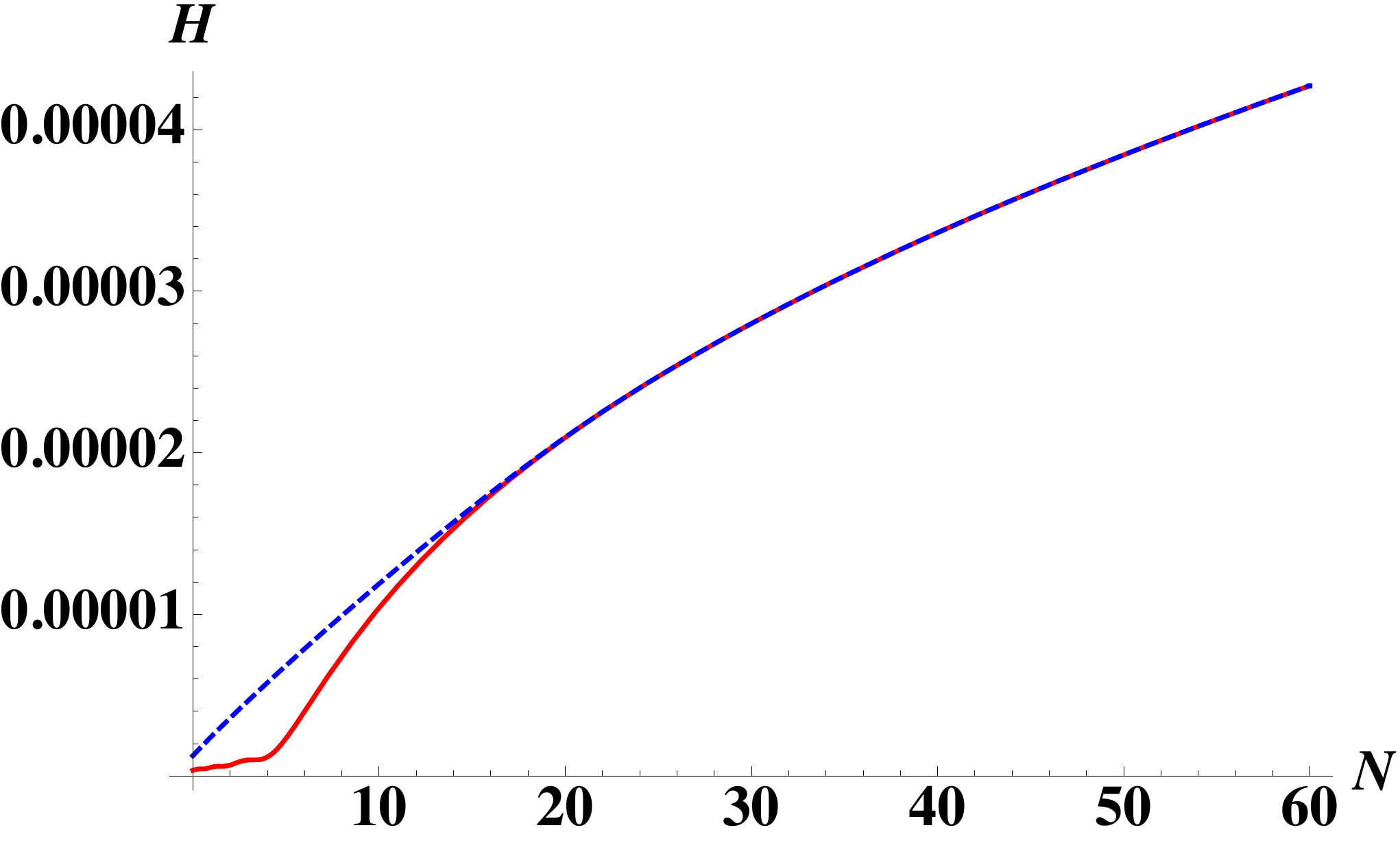}
\end{center}
\caption{\footnotesize The evolution of the Hubble scale $H$ as a function of $N$ for $\xi[N=60]=2.2$. Again the dashed blue line is the result corrected for backreaction from the sources in equations \eqref{kg} and \eqref{fr}.}
\label{H}
\end{figure}

We move to the bispectrum. The produced gauge fields lead to equilateral non-Gaussianity in the CMB \cite{Barnaby:2010vf,Barnaby:2011vw}
\be
f_{\rm NL}=\frac{\Delta^6_{\zeta}(k)}{\Delta^4_{\zeta,{\rm sr}}(k)}\, e^{6\pi\xi}\, f_{3}(\xi), \label{fnl}
\ee
where $f_{3}(\xi)$ another function defined in \cite{Barnaby:2010vf,Barnaby:2011vw}, which can be computed numerically (see \eqref{f3def} for a useful approximation). The amount of non-Gaussianity, therefore, depends exponentially on $\xi$. Between $\xi=0$ and $\xi=3$ it grows from $\mathcal{O}(1)$ to $\mathcal{O}(10^4)$ and most of the growth takes place in a small interval around $\xi\simeq 2.5$. 

The analysis of \cite{Meerburg:2012id} showed that the bounds coming from the power spectrum (especially from WMAP plus ACT, because of the violation of scale invariance) and from the bispectrum (from WMAP) are compatible, with the former being typically slightly more stringent. Spe\-ci\-fy\-ing a confidence region in $\xi$ requires assuming some prior for this parameter. The physically best motivated prior is log-flat in $\xi$ reflecting the fact the scale of the dimension five coupling $\chi F \tilde F$ could be anywhere (with strong indications that it should be below the Planck scale \cite{Banks:2003sx}). In this case at $95\%$ CL one finds $\xi\lesssim 2.2$. A flat prior on $\xi$ leads to $\xi\lesssim 2.4$.


\section{Very small scales: strong backreaction}\label{bk}

In this section we want to estimate the power spectrum and bispectrum towards the end of inflation, i.e. on scales that are too small to be observed in the CMB. The only observational handle available in this regime is the non-detection of primordial black holes, which puts an upper bound on the power spectrum \cite{malik, Carr:2009jm,Byrnes:2012yx,Klimai:2012sf,Lyth:2012yp,Shandera:2012ke}.

To make these estimates it is essential to recognize that many of the formulae described in the previous section and given in the literature about inverse decay are valid only in the regime in which backreaction on the inhomogeneous equation for $\delta\chi$ is small (see \eqref{br2}). As we show in the following, the scales relevant for the production of primordial black holes leave the horizon when backreaction is large. The authors of \cite{ng} did not account for backreaction and therefore their conclusion that gauge field production during inflation leads to black hole production might be premature.

For concreteness, we will consider a quadratic potential $V(\chi)=\frac{1}{2} m^2 \chi^2$, with the mass chosen such that at the pivot scale $k_\star$ (that we take to correspond with $N=60$) we get $\Delta^2_{\zeta}(k_\star)=2.43\cdot 10^{-9} $. 

Let us first look at the dynamics of $\chi$ and $H$. As we already discussed, when enough gauge field quanta have been produced, the conditions in \eqref{br} stop to hold (the inequality for $\ex{\edb}$ is violated first) and $\chi$ and $H$ are slowed down. As a result, inflation lasts longer. Let us check this. The behavior of $\chi$, $H$ and $\xi$ as functions of $N$ (the number of e-folds left to the end of inflation) follows from simultaneously solving \eqref{xi}, \eqref{kg} and \eqref{fr}. In Figures \ref{chi} and \ref{H} we have plotted the solutions for $\chi(N)$ and $H(N)$, with and without backreaction taken into account.  For $\xi(N=60)=2.2$, the effect of backreaction becomes $10\%$ around $N=11$.

\begin{figure}[t!]
\begin{center}
\includegraphics[scale=0.35]{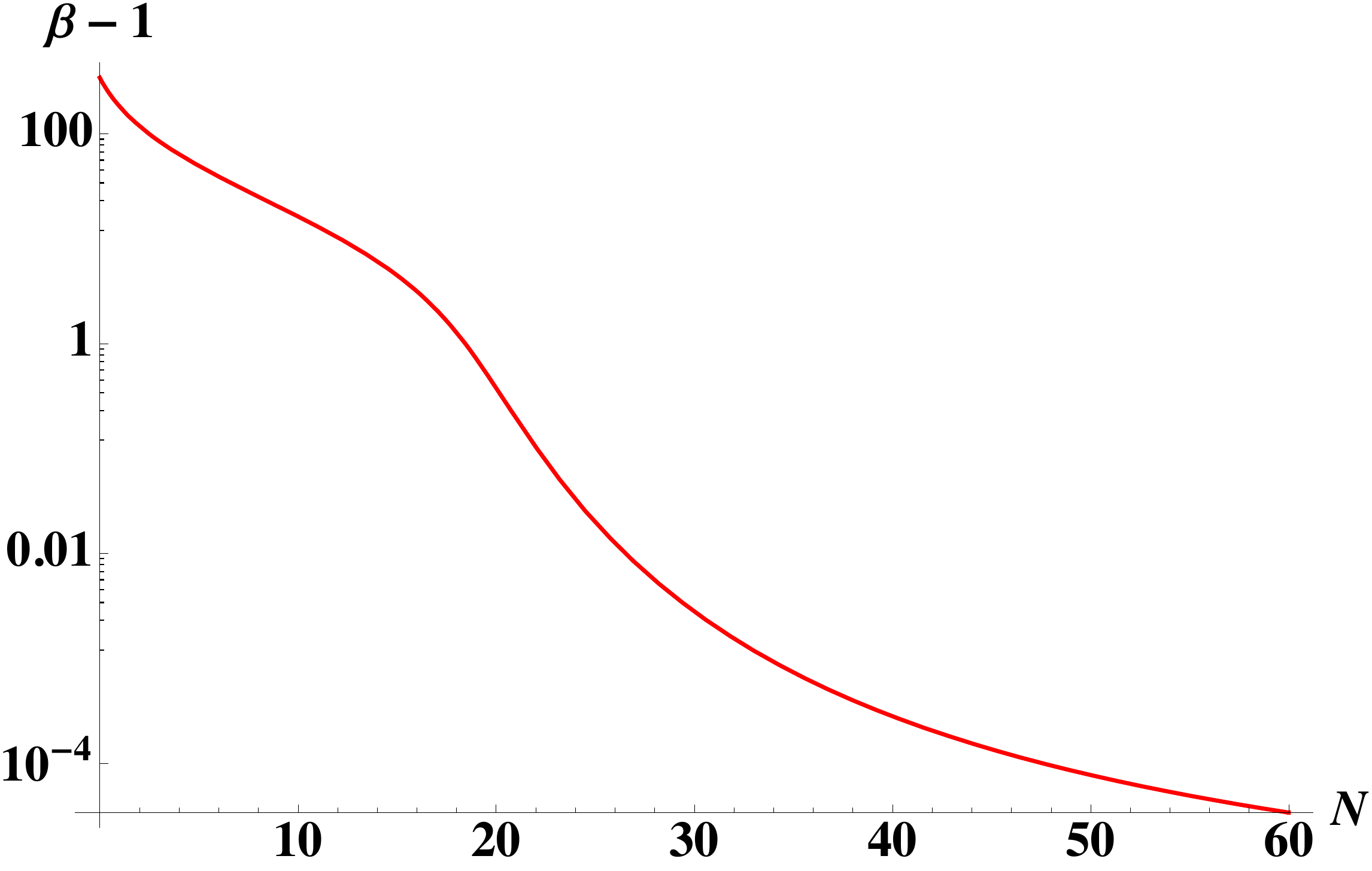}
\end{center}
\caption{\footnotesize Evolution of $(\beta-1)$ as function of $N$, for $\xi(N=60)=2.2$.}
\label{beta}
\end{figure}

Now let us consider perturbations. Of course they will be affected by the backreaction on the homogeneous dynamics $\chi$ and $H$ that we described above, but there is more. Let us consider \eqref{eom1}-\eqref{eom3}. In the last section we solved for $A$ in a \textit{homogeneous} background and used that result \eqref{asol} to compute the source term in the equation for $\chi$ perturbations. But as $ A$ and $\delta \chi$ grow larger toward the end of inflation (both of them grow as $e^{2\pi \xi}$) the source in the right-hand side of \eqref{eom1} can not be neglected anymore. If we were able to solve this equation, we would find that $\edb$ now depends on the perturbation $\delta \chi$. Expanding $\edb$, which is the source term in \eqref{eom2}, in powers of $\delta\chi$ we would find several new terms including additional friction and a modified speed of sound. In \cite{Anber:2009ua, Barnaby:2011qe} it was proposed how to estimate these effects in the regime of strong backreaction by just considering the additional friction term $\dot{\delta\chi}$. The equation of motion for the perturbation $\delta \chi$ becomes 
\ba
\ddot{\delta\chi} + 3 \beta H \dot{\delta\chi}-\frac{\nabla^2}{a^2}\delta\chi+\frac{\partial^2 V}{\partial\chi^2} \delta \chi &=& \alpha \Bigl[ \vec{E} \cdot \vec{B} -\ex{\vec{E} \cdot \vec{B}}\Bigr], \label{brps} \nonumber \\ \label{dchieq}
&&
\ea
with the additional friction term
\be
\beta\equiv 1-2\pi \xi \alpha \frac{\ex{\vec{E} \cdot \vec{B}}}{3 H \dot{\chi}}. \label{defbeta}
\ee
Here the new term in $\beta$ is caused by the dependence of $\ex{\edb}$ on $\dot{\chi}$ (via its dependence on $\xi$). The behavior of $\beta$ has been plotted in Figure \ref{beta}. It is always positive\footnote{We work with negative $\dot{\chi}$ which yields positive $\ex{\edb}$, while working with $\dot{\chi}>0$ gives $\ex{\edb}<0$.}.

The new source of backreaction can be neglected as long as
\be
2\pi \xi \alpha\frac{\ex{\vec{E} \cdot \vec{B}}}{3H\dot{\chi}}\ll1. \label{br2}
\ee
Note (from comparison with \eqref{br}) that the factor of $2\pi \xi$ makes that backreaction on the power spectrum will become significant before backreaction on $H$ and $\chi$ does. For $\xi(N=60)=2.2$ we find that backreaction becomes of order 10\% ($\beta=1.1$) at $N=22$.

\begin{figure}[t!]
\begin{center}
\includegraphics[scale=0.4]{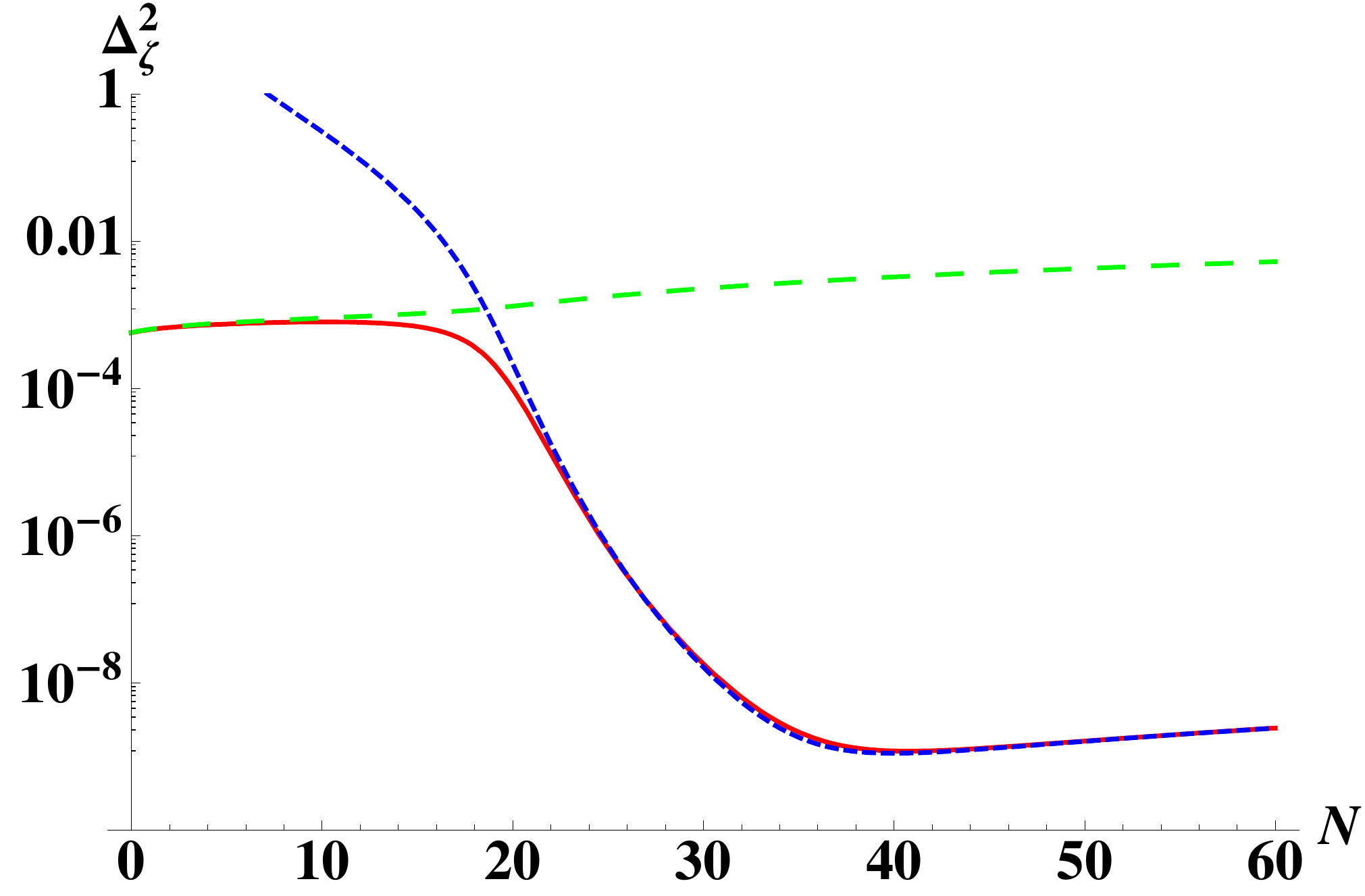}
\end{center}
\caption{\footnotesize Evolution of the power spectrum as function of $N$, for $\xi(N=60)=2.2$. The expression \eqref{spectrum} that does not take backreaction into account is in tinily dashed blue. In solid red is the estimate \eqref{psest}. When backreaction becomes significant this estimate coincides with the late-time estimate $(2 \pi \xi[N])^{-2}$, in largely dashed green.}
\label{ps}
\end{figure}

The modified equation of motion (\ref{dchieq}) suggests that (as was already noted in \cite{Barnaby:2011qe}, see also appendix \ref{turbops}) we can estimate
\be
\delta \chi \approx \frac{\alpha\left(  \edb-\ex{\edb} \right) }{3\beta H^2}
\ee
which leads to the power spectrum 
\be
\Delta_\zeta^2(k) \simeq \ex{\zeta(x)^2} \simeq \left( \frac{\alpha \ex{\edb}}{3\beta H \dot{\chi}} \right)^2. \label{psest}
\ee
This estimate turns out to be particularly good in the regime in which we can check it, i.e. when $\xi\lesssim 4$ when the backreaction is negligible and we can compare with \eqref{spectrum} (see appendix \ref{turbops}). This gives us confidence to use it also in the strong backreaction regime. It is easy to see that when backreaction becomes large, the second term in \eqref{defbeta} dominates, and we end up with
\be
\Delta_\zeta^2(k) \simeq  \left( \frac{1}{2\pi\xi} \right)^2. \label{psest2}
\ee
The estimate \eqref{psest} for the power spectrum has been plotted in Figure \ref{ps} together with the formula \eqref{spectrum}, valid only when backreaction is negligible. Indeed, in the regime of strong backreaction the power spectrum asymptotes the estimate in \eqref{psest2}. At the end of inflation we have $\xi \simeq 6.7$ (for $\xi(N=60)=2.2$), which gives 
\be
\Delta_\zeta^2(k)\simeq 7.5\cdot 10^{-4}. \label{finalps}
\ee

\section{Bounds from primordial black holes}\label{BH}

Now let us try to compare this with the existing bounds on the power spectrum coming from the non-detection of primordial black holes. These will form if at horizon re-entry (i.e. smoothing $\zeta$ on scales of order $H$) we have $\zeta>\zeta_c$, with $\zeta_c\sim 1$ denoting the critical value leading to black hole formation. If one assumes that $\zeta$ follows a Gaussian distribution (with $\ex{\zeta}=0$) one can express the probability of having $\zeta>\zeta_c$ in terms of the variance $\ex{\zeta^2}$ by analyzing the Gaussian probability distribution function. This probability corresponds to the fraction of space $b$ that can collapse to form horizon-sized black holes. Hawking evaporation and present day gravitational effects constrain this fraction $b$. Typically one finds $b$ in the range $\left(10^{-28}-10^{-5}\right)$, with the strongest bounds coming from CMB anisotropies \cite{Carr:2009jm} (spectral distortion and photodissociation of deuterium lead to a bound $b\lesssim 10^{-20}$, as for example in \cite{malik}). Setting $\zeta_c=1$ gives for the upper bound on the power spectrum \cite{Lyth:2012yp} 
\be
\Delta_{\zeta,c}^{2}(k) \simeq \ex{\zeta(x)^2} \simeq 0.008-0.05.
\ee
Here the lower bound corresponds to $b=10^{-28}$ and the upper bound to $b=10^{-5}$.

However, in our case $\zeta$ does not follow a Gaussian distribution. Instead we have (see Appendix \ref{turbops})
\be
\zeta=-\frac{\alpha\left(  \edb-\ex{\edb} \right) }{3\beta H \dot{\chi}}.
\ee
The stochastic properties of the vector field $ A$ are close to those in a free theory, i.e. it has Gaussian perturbations around $\ex{ A}=0$. 
As a consequence we can write \footnote{Here we can safely neglect the linear term, which is just the standard vacuum slow-roll contribution to $\zeta$. See also our estimate for $f_{\rm NL}$ at small scales in Appendix \ref{biest}.}
\be
\zeta = g^2-\ex{g^2} \label{ngapp}
\ee
with $g$ a Gaussian distributed field. This model was studied in \cite{Lyth:2012yp} and we follow that derivation (see also \cite{Byrnes:2012yx,Klimai:2012sf}). The probability distribution function of $\zeta$ follows from setting $P(\zeta) d\zeta = P(g) dg$, and takes the form
\be
P(\zeta)=\frac{1}{\sqrt{2\pi (\zeta+\sigma^2)}\sigma} e^{-\frac{\zeta+\sigma^2}{2\sigma^2}},
\ee
with $\sigma^2\equiv\ex{g^2}$. For a given value of $b$ we can again infer $\sigma^2$. Setting $t\equiv \frac{\zeta}{\sigma^2}+1$ (and $t_c\equiv \frac{\zeta_c}{\sigma^2}+1$) we have $d\zeta =\sigma^2 dt$ which gives
\be
b = \int_{\zeta_c}^\infty P(\zeta) d\zeta 
= \int_{t_c}^\infty  \frac{e^{-\frac{t}{2}}}{\sqrt{2 \pi  t}}  d t
=&{\rm Erfc}\left(\sqrt{\frac{t_c}{2}}\right), \label{betaps}
\ee
where ${\rm Erfc}(x)\equiv 1-{\rm Erf}(x)$ is the complementary error function. Taking again $b$ in the range $10^{-28} - 10^{-5}$ one gets a tighter upper bound on the power spectrum than in the Gaussian case:\footnote{A similar, but less precise estimate was made in \cite{Dimopoulos:2012av}}
\ba
\Delta_{\zeta,c}^{2}(k) &\simeq& \ex{\zeta(x)^2}=2\ex{g^2}^2\nn\\
& \simeq &1.3\cdot 10^{-4} - 5.8\cdot 10^{-3}. \label{finbound}
\ea

Now let us estimate what value of $b$ is relevant for our investigation. 

At the end of inflation, the total mass concentrated in the volume associated with perturbations leaving the horizon $N$ e-foldings before the end of inflation with the Hubble constant $H$ can be estimated by
\be \label{MN}
M_{N}\simeq \frac{4}{3}\pi \rho r^{3}\simeq \frac{4 \pi \Mpl^{2}}{H}e^{3N}\,,\label{MBH1}
\ee
where we reinserted the reduced Planck mass $\Mpl$, which was set to one in the rest of the paper, and $H$ is calculated at the end of inflation.  In order to study the subsequent evolution of matter in the comoving volume corresponding to this part of the universe, one should distinguish between two specific possibilities depending on the dynamics of reheating after inflation, discussed in section \ref{reh}.

If reheating is not very efficient, then the universe for a long time remains dominated by scalar field oscillations, with the average equation of state $p=0$. In this case, the total mass in the comoving volume does not change, and therefore at the moment when the black hole forms, its mass $M_{BH}$ is equal to $M_{N}$ evaluated in (\ref{MN}). For the parameters of our model, this gives an estimate (see appendix \ref{a:bh} for details)
\be \label{MN2}
M_{BH}\simeq 10\, e^{3N}~ {\rm g}\,.\label{MBH2}
\ee
On the other hand, if reheating is efficient, then the post-inflationary universe is populated by ultra relativistic particles and the energy density in comoving volume scales inversely proportional to the expansion of the universe. In this case, the black hole mass can be estimated as (see appendix \ref{a:bh})
\be \label{MN3}
M_{BH}\simeq 10\, e^{2N}~ {\rm g}\,.\label{enrbh}
\ee
In our estimates of the black hole production we will assume the latter possibility, though in general one may have a sequence of the first and the second regime. The final conclusion will only mildly depend on the choice between these two possibilities.

\begin{figure}[t!]
\begin{center}
\includegraphics[scale=0.35]{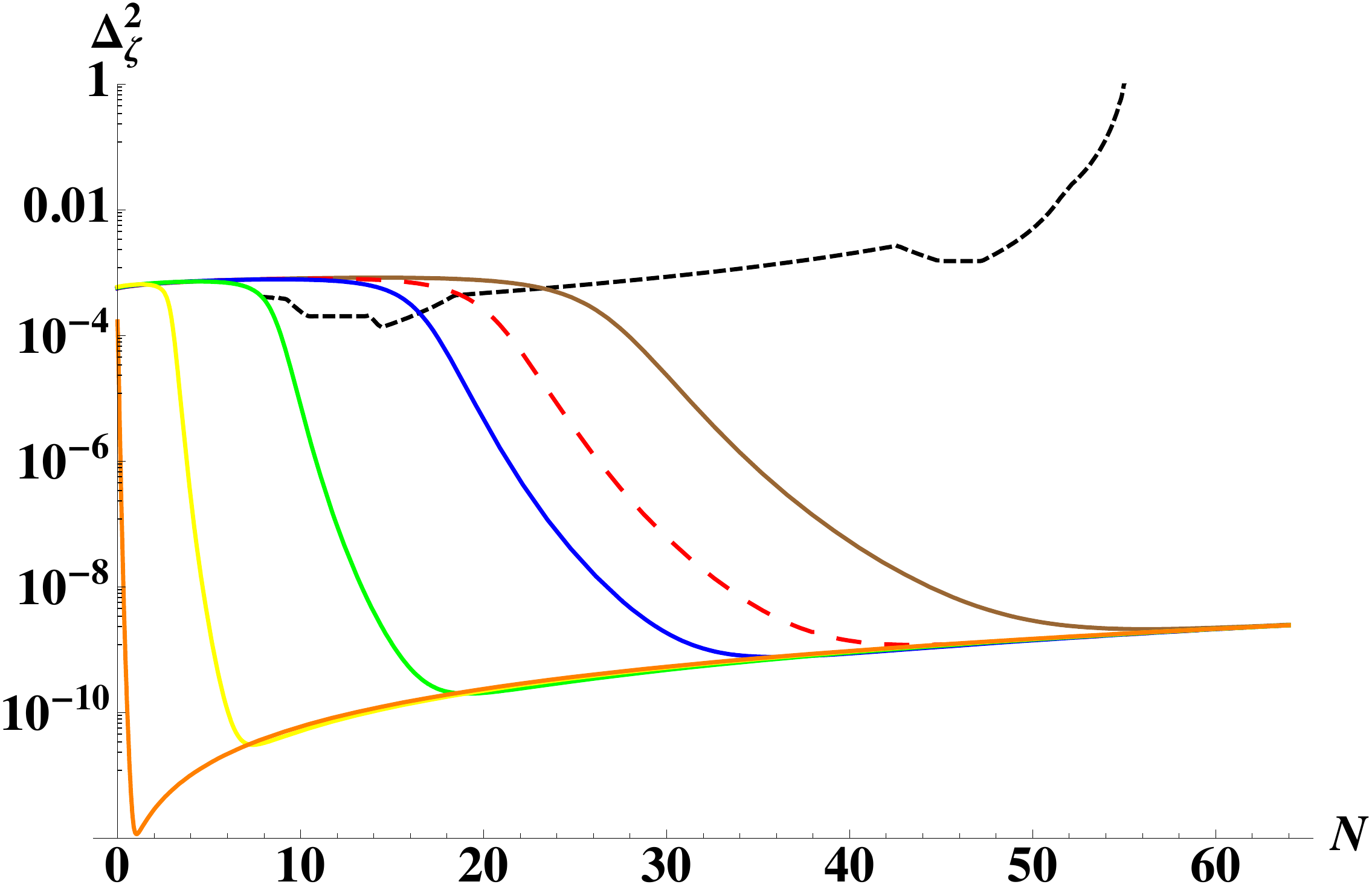}
\end{center}
\caption{\footnotesize Evolution of our estimate for the power spectrum as a function of $N$. In dashed red is the result for $\xi[N=64]=2.2$. Other lines are for $\xi[N=64]=2.5$ (solid brown), $\xi[N=64]=2$ (solid blue), $\xi[N=64]=1.5$ (solid green), $\xi[N=64]=1$ (solid yellow) and $\xi[N=64]=0.5$ (solid orange). The black hole bound is in dashed black.}
\label{bhvarxi}
\end{figure}

Now, the bounds on $b$ in terms of the would-be black hole mass $M_{BH}$ were given in \cite{malik} and updated in \cite{Carr:2009jm}. Here we follow the result in \cite{Carr:2009jm}.\footnote{However, we do not take the constraints for $M_{BH}<10^8$ g into account, as these are either very model dependent, or assume that black hole evaporation leaves stable relics.} Using \eqref{betaps} and our estimates of the black hole mass as a function of $N$, we can translate this into a bound on the power spectrum as a function of $N$. The result is in Figure \ref{bhvarxi}.

Our estimate \eqref{finalps} \textit{violates this bound} for all $N\lesssim20$ by a factor of about six. Since we have made some approximations both in deriving the late time power spectrum and in deriving its observational upper bound, our estimate could well be off by some order one factor and therefore we can not draw a definitive conclusion. It is clear though that the parameter values giving rise to an observable but not yet ruled out violation of scale invariance and non-Gaussianity in the CMB-window produce a late time power spectrum that comes at least very close to the primordial black hole bound. A more precise computation is needed to establish whether this bound is actually violated or not.

However, if such a computation revealed that primordial black holes do indeed constrain these models, that would yield a much stronger bound on $\xi$ as the ones coming from non-Gaussianity and the violation of scale invariance. Since we have seen that the power spectrum has a late-time asymptotic of $\left(2\pi \xi[N]\right)^{-1}$, this problem persists on a wide range of values for $\xi$.

For all values of $\xi$, our estimate for the power spectrum sharply increases before the end of inflation, the closer to the end the smaller $\xi$ is. However, if we disregard black hole bounds for $M_{BH}\lesssim 10^8$ g, which rely on uncertain model dependent assumptions, there are no black hole bounds for $N\lesssim 8$. From figure \ref{bhvarxi} we then see that we get
\be
\xi (N_{CMB}) \lesssim 1.5
\ee
for the bound on $\xi$ at CMB scales from primordial black hole production. In terms of the coupling constant $\alpha$, this bound implies the constraint \be
\alpha \lesssim 23.
\ee
This bound is derived using \eqref{enrbh}, i.e.~radiation domination right after the end of inflation. This assumption fixes the expansion history of the universe and therefore specifies $N_{CMB}\simeq 64$, for the $N$ corresponding to CMB scales (see appendix \ref{a:bh} for a derivation). This is required for consistency but changes the numerics very little. Therefore in all other sections we still use $N_{CMB}=60$.

For the matter domination regime, the black hole masses would be greater, for a given $N$, see (\ref{MBH2}), and therefore we would have a slightly stronger constraint on $\xi$ and $\alpha$. We find $\xi\lesssim1.3$ which corresponds to $\alpha \lesssim 20$. Instead of concentrating on it, we will now investigate the model where non-Gaussian perturbations may be generated for much smaller $\xi$ and $\alpha$, without leading to the primordial black hole problem. 


\section{Local non-Gaussianity from heavy vector fields}\label{mA}

Now let us turn to a scenario, described in \cite{Meerburg:2012id},  in which the produced gauge fields are massive. The production of gauge quanta decreases with the mass of the gauge fields: for $m_A \sim \xi H$ all production is killed. In this scenario, the gauge fields get their mass via the Higgs mechanism. Fluctuations in the Higgs field $h$ lead to fluctuations in $m_A$, which in turn generate fluctuations in the amount of produced gauge quanta, and therefore in the amount of extra friction in the dynamics of $\chi$ and $H$. In the end, one has perturbations in $\Delta N$, namely the number of extra e-folds of inflation due to gauge field production. This leads to a non-Gaussian signal in the CMB of the local type \cite{Meerburg:2012id}. Using the $\delta N$ formalism one finds
\be \label{Enrico}
f^{{\rm local}}_{\rm NL}\sim 10^{2} \left(\frac{\Delta N_{\rm max}^{3/4}\,e}{\xi\, 10^{-3}}\right)^{4}\, \left(\frac{m_{A}}{\xi H}\right)^{2}\,.
\ee
Here $\Delta N_{\rm max}$ is the increase of the duration of inflation for the case where the vector fields are massless, $h$ is the Higgs-like field responsible for the mass of the gauge field, $e$ is its $U(1)$ charge, $m_{A} = e h$ and we assumed a quadratic inflaton potential, so that $H = {m\chi\over\sqrt 6}$.

For a complete description we refer the reader to the original reference \cite{Meerburg:2012id}, section 7. Here we only want to stress that this scenario can also work for $\xi \sim 1$. Then it will surely satisfy the bounds from primordial black holes.  

Note that the classical field $h(x)$, which gives the vector field mass $e h$, can be produced either due to the tachyonic mass of the field $h$ at $h = 0$, as in the standard Higgs model, or due to accumulation of long wavelength inflationary perturbations of the field $h$. In both cases, the mechanism of \cite{Meerburg:2012id} requires that the mass of the field $h$ during inflation should be smaller than the Hubble constant. As a result, even if one assumes that the field has the standard Higgs potential, the value of the field during inflation does not correspond to the position of the minimum of the potential. Instead of that, the field takes different values in different exponentially large parts of the universe. The value of 
$f^{{\rm local}}_{\rm NL}$ in this scenario will depend on a typical local value of the field $h$, which can be determined by the stochastic approach to investigation of curvaton fluctuations \cite{Demozzi:2010aj}. 

For simplicity, and to make a clear link to the investigation performed in \cite{Demozzi:2010aj}, we will call the light field $h$ the curvaton, but one should remember that the mechanism of conversion of perturbations of the curvaton field to adiabatic perturbations is different, involving a complicated dynamical processes during reheating. In our case, fluctuations of the field $h$ lead to fluctuations $\delta N$ during inflation, and thus to a direct production of adiabatic perturbations of metric.

This scenario can work only if we have a charged scalar field with mass much smaller than $H$. At the first glance, one could achieve it by assuming that the relatively light field $S$ plays the role of the Higgs field. However, the superpotential $W=m S\Phi$ would break gauge invariance unless we assume that the field $\Phi$ is also charged. This would be inconsistent with the postulated functional form of the K{\"a}hler\, potential.\footnote{We are grateful to the referee for attracting our attention to this issue.} Therefore we must add to our model at least one charged scalar field $Q$.

Fortunately, one can easily do it. Just like in the simplest supersymmetric version of the abelian scalar electrodynamics, one should consider the charged field $Q$ without any superpotential associated with it. In the global SUSY limit, the simplest version of this theory with vanishing FI coefficient would contain D-term potential $V_{D} = {g^{2}\over 2} (\bar Q Q)^{2}$, but it would not induce any mass of the field $Q$.

However, in supergravity, the radial component $h/\sqrt 2$ of the scalar field $Q$ does acquire mass, depending on the choice of the K{\"a}hler\, potential. (The complex phase of the field $Q = {h\over \sqrt 2} e^{{i\theta}}$ is eliminated due to the Higgs effect.) We will consider the following addition to the K{\"a}hler\, potential of our model:
\be
\Delta K = Q\bar Q + \kappa\, Q\bar Q S\bar S
\ee
Terms of similar functional form were included in many versions of our inflationary scenario for stabilization of the inflaton trajectory. One can easily find that the resulting mass squared of the field $h$ during inflation is given by
\be
m^{2}_{h} = 3H^{2}(1-\kappa) \ .
\ee
Thus in the absence of the term $\kappa Q\bar Q S\bar S$ the field $h$ would be too heavy, but by considering models with $\gamma \equiv 3(1-\kappa) \ll 1$ one can have a consistent theory of a light charged scalar field with mass squared $\gamma H^{2}$ with $\gamma \ll 1$, as required. 

Of course, this requires fine-tuning, but this is just a price which one should be prepared to pay for the description of non-Gaussian inflationary perturbations.  We will study observational consequences of this model
in the next section.

\section{Stochastic approach}

\label{stoch}

In this section we want to find out how fluctuations in the curvaton field $h$ lead to a variable gauge field mass, and therefore to a non-Gaussian signal in the CMB.
We will begin our study with investigation of the behavior of the
distribution of the fluctuations in $h$, following \cite{Demozzi:2010aj}. 
During inflation, the long-wavelength distribution of this field generated
at the early stages of inflation behaves as a nearly homogeneous classical
field, which satisfies the equation%
\begin{equation}
3H\dot{h}+V_{h }=0\ ,  \label{a1}
\end{equation}%
which can be also written as
\begin{equation}
{\frac{dh ^{2}}{dt}}=-{\frac{2V_{h }\,h }{3H}}\ .
\label{a3ax011}
\end{equation}%
However, each time interval $H^{{-1}}$ new fluctuations of the scalar field
are generated, with an average amplitude squared\footnote{For a real massless field we would get ${\langle \delta h^{2}\rangle }={\frac{H^{2}}{4\pi ^{2}}}$. An extra coefficient $2$ appears in  (\ref{a11}) because the field $Q$ is complex, so its absolute value changes faster because of independent fluctuations of its two components. One could argue that in the unitary gauge we only have one scalar degree of freedom. However, unitary gauge is problematic in the description of the Brownian motion and cosmic string formation in the Higgs model. We present the results which should be valid in the regime of small gauge coupling constant $e$. Our main conclusions are unaffected by this factor of 2 issue.}
\begin{equation}
{\langle \delta h ^{2}\rangle }={\frac{H^{2}}{2\pi ^{2}}}\ .
\label{a11}
\end{equation}%

The wavelength of these fluctuations is rapidly stretched by inflation. This
effect increases the average value of the squared of the classical field $%
h $ in a process similar to the Brownian motion. As a result, the
square of the field $h $ at any given point with an account taken of
inflationary fluctuations changes, in average, with the speed which differs
from the predictions of the classical equation of motion by ${\frac{H^{3}}{%
4\pi ^{2}}}$: 
\begin{equation}
{\frac{dh ^{2}}{dt}}=-{\frac{2V_{h }\,h }{3H}}+{\frac{H^{3}}{%
2\pi ^{2}}}\ .  \label{a3ax0}
\end{equation}%

Using $3H\dot{\chi}=-V_{\chi}$, one can rewrite this equation as 
\begin{equation}
{\frac{dh ^{2}}{d\chi}}={\frac{2V_{h }\,h }{V_{\chi}}}-{%
\frac{V^{2}}{6\pi ^{2}V_{\chi}}}\ .  \label{a3ax}
\end{equation}%

By solving this equation with different boundary conditions, one can find the most probable value of the locally homogeneous field $h$.

Now we will consider the case when the mass of the curvaton field is given
by 
\begin{equation}
m_{h }^{2}=\gamma H^{2}= {\frac{\gamma m^{2}\chi ^{2}}{6}}
\end{equation}%
with $\gamma \ll 1$.
This corresponds to the total potential 
\begin{equation}
V(\chi,h)= {m^{2}\over 2} \chi^{2} + {\gamma\over 2} H^{2}h^{2}.  \label{vchiH}
\end{equation}%
We assume that $h \ll 1$, and therefore one can  estimate $H^{2}\approx {m^{2}\over 6} \chi^{2}$.
In this case, Eq. (\ref{a3ax}) becomes 
\begin{equation}
{\frac{dh ^{2}}{d\chi}}={\frac{2\gamma \chi h^{2}}{6}}-{\frac{m^{2}\chi^{3}}{24\pi ^{2}}}\ .  \label{a3ax1}
\end{equation}%

This equation has a family of different solutions, 
\begin{equation}
h ^{2}={3m^{2}\over 4\pi^{2}\gamma^{2}} \left(1+\gamma{\chi^{2}\over 6}\right)  + A\ e^{\gamma\chi^{2}/6},  \label{a3ax2}
\end{equation}
where $A$ is a constant which could be either positive or negative, depending on initial conditions.
During inflation these solutions converge to a simple attractor solution
\begin{equation}
h= {\sqrt 3 m\over 2\gamma \pi} \sqrt{1+ {\gamma\chi^{2}\over 6}}\ .  \label{a3ax3}
\end{equation}
We are interested in using this formula to estimate the size of non-Gaussianity, which is produced by the conversion of perturbations in $h$ into curvature perturbations when the backreaction from gauge fields on the homogeneous evolution becomes substantial, i.e.~close to the end of inflation. Hence we should take $\chi\sim 1$ in (\ref{a3ax3}). For $\gamma\ll 1$, this solution approaches a constant $h= {\sqrt 3 m\over 2\gamma \pi}$ during the last stages of inflation. Note that this a posteriori justifies the assumption that $h \ll 1$, as long as $\gamma\gg 10^{-6}$.

To give a particular numerical example, we will use (\ref{Enrico}) for the case $\xi = 0.5$. A numerical analysis shows that in this case $\Delta N_{\rm max} \sim 0.044$, and therefore
\be
f^{{\rm local}}_{\rm NL}\sim 2.5\times 10^{11}  e^{6}\gamma^{{-2}}\chi^{-2} \, 
\ee
at the end of inflation with ${\gamma\chi^{2}/ 6}\ll 1$.

All our approximations should work fine if the mass of the vector field is much smaller than $H$, which leads to a constraint $e\ll \gamma\chi$.

Consider for example $\gamma = 0.1$ and $\chi = 1$, which corresponds to the very end of inflation. (We should stress that it would not be consistent to take $\chi$ much larger than $\mathcal{O}(1)$ in Planck units since that is its value when curvature perturbations are generated in our scenario. Moreover,  the main contribution to $\Delta N_{\rm max}$ is given by the last part of the inflationary trajectory where $\chi = \mathcal{O}(1)$.) In this case
\be
f^{{\rm local}}_{\rm NL}\sim 2.5\times 10^{13}  e^{6} \, 
\ee
To have non-Gaussian perturbations with $f^{{\rm local}}_{\rm NL} = O(10)$ one should take $e\approx 10^{{-2}}$.


\section{Gauge field production in SUGRA inflation: reheating}\label{reh}
We have found that the coupling $\chi F \tilde{F}$ needed for reheating in (the pseudo scalar variant of) the new class of SUGRA inflation models proposed in \cite{Kallosh:2010ug,Kallosh:2010xz,Kallosh:2011qk} can as well yield an observable non-Gaussian signal. It only remains to be seen what the effects of the typically needed values for $\xi$ are for the reheating in the combined model.

In \cite{Kallosh:2011qk} the reheating temperature $T_R$ for the decay of a scalar inflaton field to two photons due to the interaction ${\alpha\over 4}\phi F^{2}$ was estimated as
\be
T_R \approx \sqrt{2}{\alpha} \times 10^9 {\rm GeV}. 
\ee
A similar estimate is valid in our case. One may also represent it in an equivalent way using relation
$\frac{\alpha}{4} =  -\frac{\xi H}{2\dot{\chi}}$, and an expression for the slow-roll parameter $\epsilon=\frac{\dot{\chi}^2}{2H^2}$
\be
T_R \approx \frac{2\xi}{\sqrt{\epsilon}}  \times 10^9 {\rm GeV}.
\ee
As long as one can describe reheating as a particle by particle decay, reheating in inflationary models of this type does not depend much on whether the inflaton field is a scalar or a pseudo scalar. In both types of models, one may consider interactions with $\alpha \ll 1$, which results in reheating temperature $T_{R} \lesssim 10^{8}$ GeV. This solves the cosmological gravitino problem for gravitino in the typical mass range $m_{3/2} \lesssim 1$ TeV.  

\begin{figure}[t!]
\begin{center}
\includegraphics[scale=0.35]{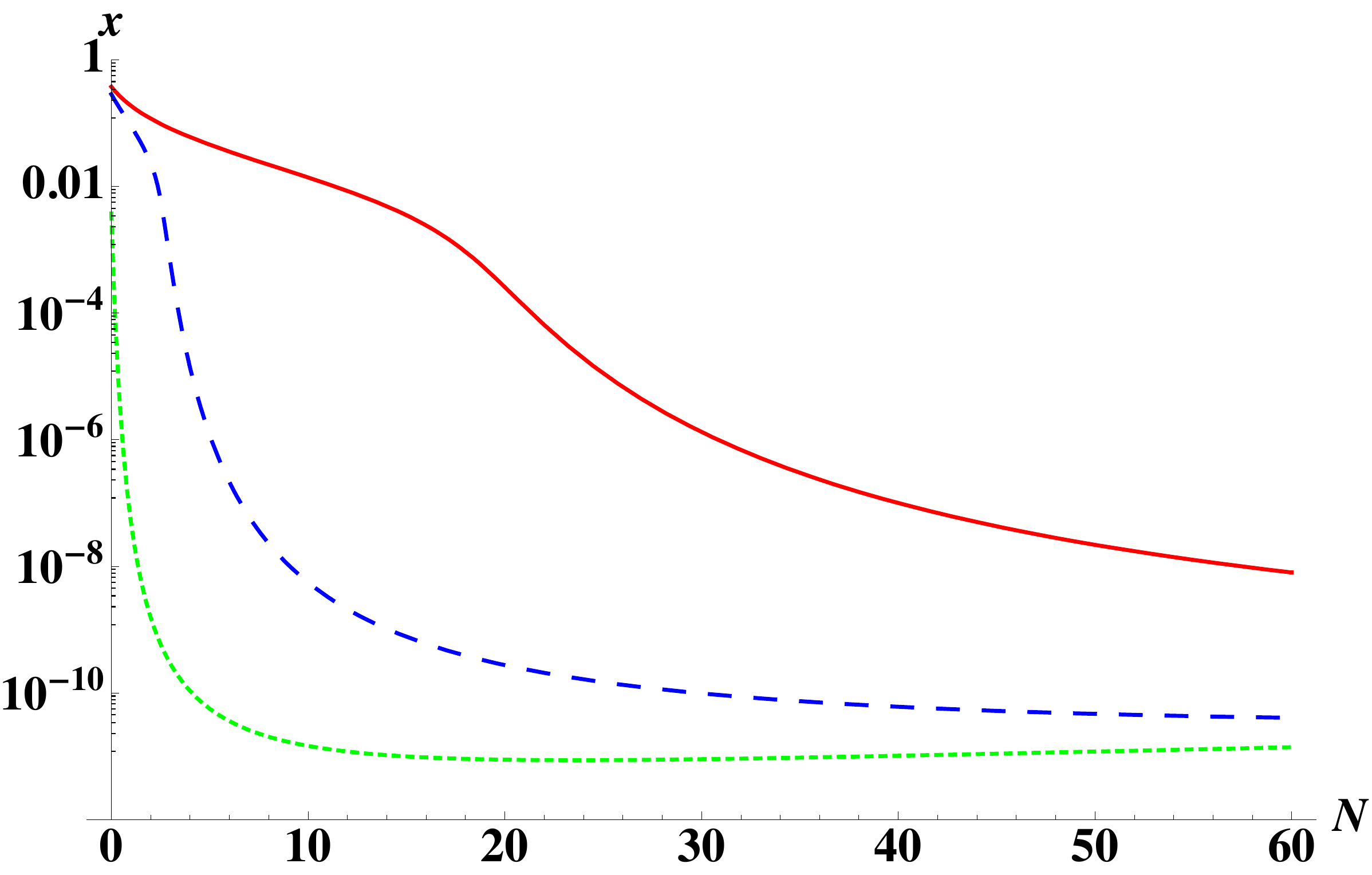}
\end{center}
\caption{\footnotesize Evolution of the normalized energy of the vector field, $x\equiv \frac{1}{2} \left( E^2 + B^2 \right) / 3H^2$ as a function of $N$, for $\xi[N=60]=2.2$ (solid red), $\xi[N=60]=1.0$ (largely dashed blue) and $\xi[N=60]=0.5$ (tinily dashed green).}
\label{reheat}
\end{figure}

However, for $\alpha \gtrsim 1$, which is required for production of non-Gaussianity in the models based on the pseudo scalar inflaton, an estimate described above gives $T_{R } > 10^{9}$ GeV. It is good for the theory of leptogenesis, but it could be bad from the point of view of the gravitino problem. Moreover, for $\alpha \gtrsim 1$ an entirely different mechanism of reheating is operating. At the end of inflation, when the time-dependent parameter $\xi$ grows and becomes large, a significant fraction of the energy of the inflaton field gradually becomes converted to the energy of the vector field (see Figure \ref{reheat}). This is a very efficient mechanism, which may lead to a very rapid thermalization of energy in the hidden sector. This may exacerbate the gravitino problem. Fortunately, this problem does not appear for  superheavy gravitino with mass $m_{3/2} \gtrsim 10^{2}$ TeV. Such gravitinos appear in many versions of the models of mini-split supersymmetry, which became quite popular during the last few years, see \cite{Dudas:2012wi,split} and references therein.

\section{Conclusions}

The new class of chaotic inflation models in supergravity needs a gauge-gauge-inflaton coupling for reheating. The inclusion of this coupling can produce gauge fields and can provide a Planck-observable but not yet ruled out non-Gaussian signal in the CMB. 

In this article we have studied two possible realizations of this scenario. Taking the parameter $\xi\simeq 2.2 - 2.5$ ($\alpha \simeq 32-37$) produces a large amount of gauge quanta, that by inverse decay give rise to an equilateral non-Gaussianity in the CMB, as studied in \cite{Barnaby:2010vf,Barnaby:2011vw}. However, we have estimated that towards the end of inflation the power spectrum grows so much that the model may be ruled out because it overproduces primordial black holes. As our order-one estimate lands within a factor of six from the critical black hole bound on the power spectrum (with the non-Gaussian nature of the signal taken into account), we need a more precise computation to draw a definitive conclusion.

In the second scenario, where the produced gauge fields are massive due to the Higgs effect in presence of a light curvaton-type field, one can take a smaller value for $\xi$, of order 0.5 - 1, corresponding to $\alpha$ from 8 to 15. Then the model is free of black hole trouble. In this case, fluctuations in the curvaton field modulate the duration of inflation and can give rise to adiabatic non-Gaussian perturbations of the local type with $f_{\rm NL}\sim \mathcal{O}(10)$. For smaller values of $\alpha$, we return to the standard chaotic inflation scenario with Gaussian adiabatic perturbations.

\section*{Acknowledgments}

A.L.~is supported by NSF grant PHY-0756174. E.~P.~is supported in part by the Department of Energy grant DE-FG02-91ER-40671. S.~M.~ wants to thank the Stanford Institute for Theoretical Physics for great hos\-pi\-ta\-li\-ty during his two month's visit in which this work was initiated, and Marieke Postma for many valuable discussions.


\appendix


\section{Variance of $\edb$}

The variance of $\edb$ is defined as
\be 
\sigma^{2}&\equiv&\langle (\edb)^{ 2} \rangle -\langle \edb \rangle^{2}\\
&=&\langle E_{i} E_{j} \rangle \langle B_{i} B_{j}\rangle + \langle E_{i} B_{j} \rangle \langle B_{i} E_{j}\rangle\,.
\ee
We find
\ba
\langle E_{i} E_{j} \rangle \langle B_{i} B_{j}\rangle&=&\frac1{a^{8}}\int \frac{dk dq}{(2\pi)^{6}} |A'(k)|^{2}|A(q)|^{2}q^{4}k^{2} \, \nn\\
&& \qquad\int d^{2}\Omega_{k} d^{2}\Omega_{q} \epsilon_{i}(k)\epsilon_{i}(q) \epsilon_{j}^{\ast}(k)\epsilon_{j}^{\ast}(q)\,,\nn\\
\langle E_{i} B_{j} \rangle \langle B_{i} E_{j}\rangle&=&\frac1{a^{8}}\int \frac{dk dq}{(2\pi)^{6}} A(k)A'^{\ast}(k) A'(q) A^{\ast}(q) \,q^{3}k^{3} \, \nn\\
&& \qquad\int d^{2}\Omega_{k} d^{2}\Omega_{q} \epsilon_{i}(k)\epsilon_{i}(q) \epsilon_{j}^{\ast}(k)\epsilon_{j}^{\ast}(q)\,.\nn\\
\ea
Here we use the polarization tensor conventions given in \cite{Barnaby:2011vw}:
\ba
\vec{k} \cdot \vec{\epsilon}_\pm(\vec{k})&=&0\nn\\
 \vec{k} \times \vec{\epsilon}_\pm(\vec{k})& =&\mp ik \vec{\epsilon}_\pm (\vec{k})\nn\\
  \vec{\epsilon}_\pm(-\vec{k}) &=& \vec{\epsilon}_\pm (\vec{k})^\star,
\ea
which are normalized via $\vec{\epsilon}_\lambda(\vec{k})^\star \cdot \vec{\epsilon}_{\lambda'}(\vec{k}) = \delta_{\lambda \lambda'}$. Given our conventions (see footnote \ref{footnote4}) we are dealing with $\vec{\epsilon}_-$ here.

The angular integral gives $(4\pi)^{2}/3$, i.e.~a third of the whole sphere. The integrals over the modulus are similar to the one in \cite{Barnaby:2011vw} and are computed in the same way
\be 
I_{2}&=&\frac1{a^{4}}\int \frac{dk }{(2\pi)^{3}} |A'(k)|^{2} k^{2}\simeq 2.2 \cdot 10^{-5} \frac{H^{4}}{\xi^{3}}e^{2\pi \xi}\,,\\
I_{3}&=&\frac1{a^{4}}\int \frac{dk }{(2\pi)^{3}}\frac{\partial_\tau}{2}|A(k)|^{2} k^{3}\simeq 1.9 \cdot 10^{-5} \frac{H^{4}}{\xi^{4}}e^{2\pi \xi}\,,\\
I_{4}&=&\frac1{a^{4}}\int \frac{dk }{(2\pi)^{3}} |A(k)|^{2} k^{4}\simeq 1.9 \cdot 10^{-5} \frac{H^{4}}{\xi^{5}}e^{2\pi \xi}\,.
\ee
Putting things together one finds
\be
\sigma&=&\sqrt{\frac{(4\pi)^{2}}{3} \left(I_{3}^{2}+I_{2}I_{4}\right)}\\
&=&2.0 \cdot 10^{-4} \frac{H^{4}}{\xi^{4}}e^{2\pi \xi}\simeq \langle \edb \rangle\,. \label{sigma}
\ee

\section{Power spectrum estimate} \label{turbops}
In \cite{Barnaby:2010vf,Barnaby:2011vw} the power spectrum \eqref{spectrum} has been obtained by the Green's function method. In \cite{Barnaby:2011qe} a quick estimate was introduced to compute the power spectrum in the case of large backreaction ($\beta\gg1$). Here we want to review and further explore this estimate, showing how it leads to \eqref{psest} and also how, in the case of negligible backreaction, it approximates the precise result \eqref{spectrum} within a factor of two. 

The full equation of motion for the perturbation $\delta \chi$ is (in real space)
\ba
\delta \ddot{ \chi} +3\beta H\delta \dot{\chi} -\frac{\nabla^2}{a^2}\delta\chi + m^2 \delta \chi&=&\alpha\left[   \vec{E}\cdot \vec{B} - \langle \vec{E}\cdot \vec{B} \rangle\right],\nn\\
&&
\ea
with
\be
\beta\equiv 1-2\pi \xi \alpha \frac{\ex{\edb}}{3 H \dot{\chi}}.   \label{defbeta2}
\ee
Near horizon crossing we can estimate $\partial \sim H$. Since we have, near horizon crossing, $H^2=\frac{k^2}{a^2}$, the first term cancels the third one. The second term can be approximated as $3\beta H^2 \delta \chi$. The last term on the left hand side is just a slow-roll correction and can be discarded. This directly gives
\be
\delta \chi \approx \frac{\alpha\left(  \edb-\ex{\edb} \right) }{3\beta H^2}
\ee
and therefore we have
\be
\zeta\equiv-\frac{H}{\dot{\chi}}\delta \chi \approx -\frac{\alpha\left(  \edb-\ex{\edb} \right) }{3\beta H \dot{\chi}}. \label{zetanong}
\ee
For the position space two point function of $\zeta$ we immediately get
\ba
\langle\zeta(x)^2\rangle\equiv \frac{H^2}{\dot{\chi}^2}<\delta \chi^2>&\approx& \frac{H^2}{\dot{\chi}^2} \left( \frac{\alpha\sigma}{3\beta H^2}\right)^2\nn\\
& = &  \left( \frac{\alpha \ex{\edb}}{3\beta H \dot{\chi}}\right)^2   \label{r}
\ea
with $\sigma$ the variance computed in the previous subsection.

 To compare the position space power spectrum with the momentum space power spectrum we use
\ba
\ex{\zeta(\vec k) \zeta(\vec k')}&\equiv&(2\pi)^{3} \delta^{3} \left(\vec k+\vec k'\right) P(k)\,,\quad  P(k)\equiv\frac{2
\pi^{2} \Delta_{\zeta}^{2}(k)}{k^{3}}\,,\nn \\
\ex{\zeta(x)^{2}}&=&\int d\ln k\,\Delta_{\zeta}^{2}(k)\simeq \mathcal{O}(1) \Delta_{\zeta}^{2}(k).
\ea
This gives the result \eqref{psest}:
\be
\Delta_\zeta^2(k) \simeq \ex{\zeta(x)^2} = \left( \frac{\alpha \ex{\edb}}{3\beta H \dot{\chi}} \right)^2. \label{finalpsest}
\ee

This expression has been plotted in Figure \ref{ps}.

Now when backreaction is strong we can approximate $\beta\approx -2 \pi \xi \alpha \frac{\ex{\edb}}{3 H \dot{\chi}}$, which immediately gives the approximation \eqref{psest2}
 \be
 \Delta_\zeta^2(k)= \frac{1}{(2\pi \xi)^2}. \label{psback}
 \ee
We can as well make an approximation for the case where $\beta\approx 1$ (negligible backreaction) and compare the result with the precise result \eqref{spectrum}, just to see how well this whole approximation works. For $\beta=1$ we have
\be
 \Delta_\zeta^2(k)=    \left(\frac{\alpha \ex{\edb}}{3 H \dot{\chi}}\right)^2.
\ee
Upon using the estimate for $\ex{\edb}$ found in \cite{Barnaby:2010vf,Barnaby:2011vw} 
\be
\ex{\edb} \approx 2.4\times 10^{-4} \frac{H^4}{\xi^4} e^{2\pi\xi} \label{edbbar}
\ee
and 
\be
\alpha \equiv -\frac{2 H \xi}{\dot{\chi}} \label{alpha}
\ee
we find
\ba
 \Delta_\zeta^2(k)&=& \frac{4 H^2 \xi^2}{\dot{\chi}^2} \times 5.76\times10^{-8}\times \frac{H^8}{\xi^8} e^{4\pi\xi}\times\frac{1}{9 H^2 \dot{\chi}^2}\nn\\
&=& 2.56\times 10^{-8} \times \frac{H^8}{\dot{\chi}^4} \times \frac{e^{4\pi\xi}}{\xi^6}\nn\\
&=& 2.56\times 10^{-8} \times \left(\frac{H^2}{2\pi\dot{\chi}}\right)^4 \times (2\pi)^4 \times \frac{e^{4\pi\xi}}{\xi^6}\nn\\
&=& 4.0 \times 10^{-5} \times \Delta_{\zeta,{\rm sr}}^4(k) \times\frac{e^{4\pi\xi}}{\xi^6}. \label{fines}
\ea

This can be compared with the more precise result computed in \cite{Barnaby:2010vf,Barnaby:2011vw} that uses the Green's function approach
\ba
 \Delta_\zeta^2(k) &=& \Delta_{\zeta,{\rm sr}}^4(k) \times f_2(\xi) \times e^{4\pi \xi} \label{dz2precise}\\
  &\simeq & \Delta_{\zeta,{\rm sr}}^4(k) \frac{7.5\times10^{-5}}{\xi^6} \times e^{4\pi\xi}.\label{f2}
\ea
where in the second line we used the large $\xi$ limit for $f_{2}$. We infer that this quick estimate is off by a factor less than two.

Actually, for some $\xi$ the estimate comes even closer than this ratio $\frac{7.5}{4}$. Let us examine the situation at $\xi=3$ (which, for $\xi(N=60)=2.2$), corresponds to $N\approx 35$). Above, we approximated the numerical function $f_2(\xi)$ by $\frac{7.5\times10^{-5}}{\xi^6}$ which yields an overestimate by a factor of 1.3. At the other hand, we also approximated the numerically found result for $\ex{\edb}$ by the estimate \eqref{edbbar}, which is an underestimate, that for $\xi=3$ only captures a fraction of 0.73 of the true $\ex{\edb}$. Putting everything together one finds that, at $\xi=3$ ($N=35$), our estimate \eqref{finalpsest} with $\beta$ set to one overestimates the precisely computed numerical result \eqref{dz2precise} by a factor of
\be
\frac{4}{7.5} \times \frac{1.3}{\left(0.73\right)^2} \approx 1.3.
\ee
At $\xi$=2.2 (N=60) we find that our estimate \ref{finalpsest} overestimates the precisely computed result by a factor of 2.5.

\begin{figure}[t!]
\begin{center}
\includegraphics[scale=0.35]{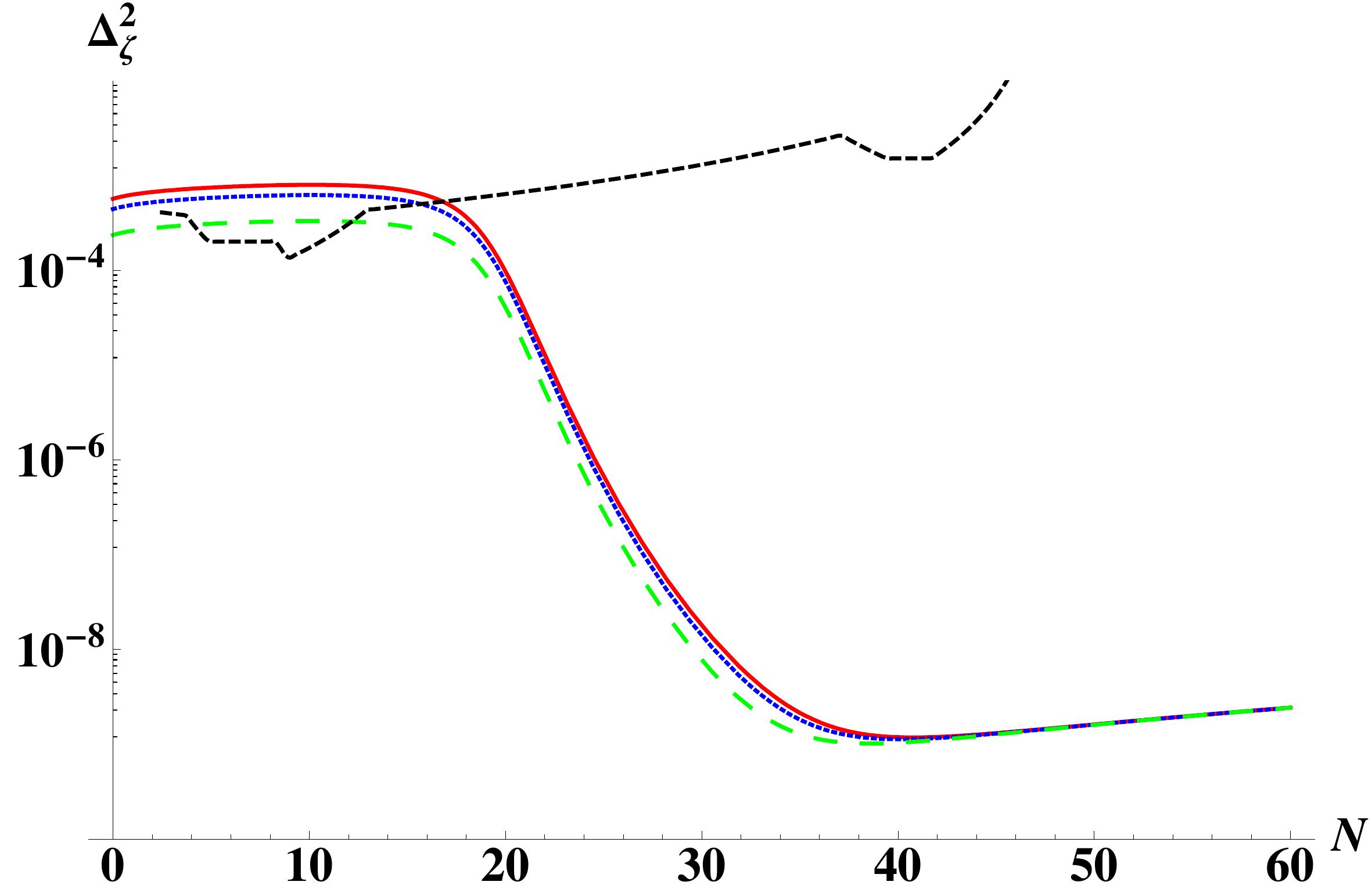}
\end{center}
\caption{\footnotesize Evolution of the power spectrum as in Figure \ref{ps}, still for $\xi[N=60]=2.2$. The red solid line is our estimate. The blue, tinily dashed line is our estimate corrected with a fudge factor of 1.3. The green, largely dashed line is our estimate corrected with a fudge factor of 2.5. All signals remain within an order one factor from the black hole bounds in dashed black.}
\label{psff}
\end{figure}

Now one might introduce a fudge factor such that at some preferred value for $\xi$ our approximation precisely matches the numerically computed result. However, we have just seen that the inclusion of such a fudge factor will induce only a small shift in our estimate that we anyway only trust up to corrections of order one. Besides, the fudge factor would always be arbitrary, as it depends on the preferred value of $\xi$ where it makes both signals match. Therefore it seems safe to neglect it altogether. In Figure \ref{psff} we have for once plotted how the total power spectrum (including the standard slow-roll contribution) would shift from such a correction. In the rest of the paper we work with our uncorrected estimate for the power spectrum.

N.B. This estimate involves only the gauge field contribution to the power spectrum. Apart from that there is always the standard slow-roll component $\Delta_{\zeta,{\rm sr}}^2(k)$. This is the dominant contribution on CMB scales. That is why any estimate of the total power spectrum matches the precise result so well on CMB scales, whatever order one fudge factor one chooses.

\section{Skewness of $\edb$}

We want to compute
\ba
\tau^{3}&\equiv &\ex{ \left(\edb-  \ex{\edb}\right)^{3}}\nn\\
&=& \ex{ \left(\edb\right)^{3}}-3\ex{\edb}\sigma^2- \ex{\edb}^{3} \nn\\
&\simeq&  \ex{ \left(\edb\right)^{3}}_{\rm c}+3 \ex{\edb}^{3}  \,, \label{taudef}
\ea
where we used that $\ex{ \left(\edb\right)^{2}}\simeq2 \ex{\edb}^{2}$ from the previous section and in the last step we recognized that there are $1+3\times 2 = 7$ non-connected diagrams in $\ex{ \left(\edb\right)^{3}}$, each one equals to $\ex{\edb}^{3} $. Using Wick's theorem we find many terms. All of them have the same angular integral
\ba
&&\int d^{2}\Omega_{k_{1}} d^{2}\Omega_{k_{2}}d^{2}\Omega_{k_{3}} \epsilon_{i}(k_{1})\epsilon_{i}(k_{2}) \epsilon_{j}^{\ast}(k_{1})\epsilon_{j}(k_{3})\,\epsilon_{j}^{\ast}(k_{2})\epsilon_{j}^{\ast}(k_{3})\nn\\
&& \qquad \qquad \qquad=\frac{2\pi^{5}}{3}\,.
\ea
Counting all the possible pairwise contractions one finds
\ba
\ex{ \left(\edb\right)^{3}}_{\rm c}&=&- \frac{2\pi^{5}}{3} \left(2I_{3}^{3}+I_{2}I_{3}I_{4}\right)\nn\\
&=&\left[-2.4 \cdot 10^{-4} \frac{H^{4}}{\xi^{4}}e^{2\pi \xi}\right]^{3}\simeq -\langle \edb \rangle^{3}\nn\\
\ea
and therefore
\be
\tau
^{3}\simeq 2 \langle \edb \rangle^{3}. \label{tau3}
\ee

\section{Bispectrum and $f_{NL}$ estimate} \label{biest}

The position space three point function of $\zeta$ can be directly generalized from \eqref{r}:
\ba
\ex{\zeta(x)^3}\equiv- \frac{H^3}{\dot{\chi}^3}\ex{\delta \chi^3}&\approx& -\frac{H^3}{\dot{\chi}^3} \left( \frac{\alpha \tau}{3\beta H^2}\right)^3\nn\\
& = & - 2\left( \frac{\alpha \ex{\edb}}{3\beta H \dot{\chi}}\right)^3 , \label{zeta3est}
\ea
where we used the definition of the skewness $h^3$ \eqref{taudef} and its estimate \eqref{tau3}. $\ex{\zeta(x)^3}$ is positive. (Again: we work with negative $\dot{\chi}$ which yields positive $\ex{\edb}$, while working with $\dot{\chi}>0$ would give $\ex{\edb}<0$.) 

Let us first analyze this result in the regime where backreaction is negligible, i.e. $\beta=1$. 
Using \eqref{edbbar} and \eqref{alpha}  we get
\ba
\ex{\zeta(\vec{x})^3} &\simeq&
2 \frac{8}{27} \left(2.4\times 10^{-4}\right)^{3}\,\frac{H^{12}e^{6\pi\xi}}{\xi^{9} \dot\chi^{6}}\nn\\
&\simeq& 8.2 \times 10^{-12} \,\frac{H^{12}e^{6\pi\xi}}{\xi^{9} \dot\chi^{6}}\,.\label{well}
\ea

Now we want to compare this with the momentum space bispectrum $B(k)$, defined via
\ba
&&\ex{\zeta(\vec k_{1})\zeta(\vec k_{2})\zeta(\vec k_{3})}\equiv (2\pi)^{3} \delta^{3} \left(k_{1}+k_{2}+k_{3}\right) B(\vec k_{1},\vec k_{2},\vec k_{3}) \nn\\
&&
\ea
for which we can write
\be
\hskip -12pt \ex{\zeta(\vec x)^{3}}=\intd{k_{1}}\intd{k_{2}} B(\vec k_{1},\vec k_{2},-\vec k_{1}-\vec k_{2})\,.
\ee
When non-Gaussianity is large mostly on equilateral triangles, the integral is supported in the region $k_{2}\simeq k_{1}$ and $\theta_{12}\simeq \pi/3$. Hence we estimate
\ba
\ex{\zeta(\vec{x})^{3}}&=& \int d\log k \frac{8 \pi^{2}}{(2\pi)^{6}}  k^{6}B_{\rm eq}(k) \nn\\
&\simeq & \frac{8 \pi^{2}}{(2\pi)^{6}}  k^{6}B_{\rm eq}(k)  \mathcal{O}(1)\,,
\ea
where $B_{\rm eq}(k)$ is the bispectrum evaluated on equilateral triangles. Now we can compare our estimate  \eqref{well} with the precisely computed result using the Green's function approach, that we take from result (2.8) of \cite{Meerburg:2012id}, 
\ba
B_{\rm eq}(k)&=& \frac{1}{(2\pi)^3} \ex{\zeta(\vec k_{1})\zeta(\vec k_{2})\zeta(\vec k_{3})} \nn\\
 &\simeq& \frac{3\times 3\times 2.8\times 10^{-7}}{10 (2\pi)^{2}} \,\frac{H^{12}e^{6\pi\xi}}{\xi^{9} \dot\phi^{6}}\frac{1}{k^6},
\ea
where we have used the large $\xi$ estimate
\be
f_3(\xi)=\frac{2.8\cdot 10^{-7}}{\xi^9}. \label{f3def}
\ee
This last result leads to
\be
\ex{\zeta(\vec{x})^{3}}\simeq\frac{8 \pi^{2}}{(2\pi)^{6}}  k^{6}B_{\rm eq}(k) \simeq 8.2 \times 10^{-12} \,\frac{H^{12}e^{6\pi\xi}}{\xi^{9} \dot\phi^{6}} \label{z3est}
\ee
which agrees (surprisingly) well with \eqref{well}.

In the regime of strong backreaction we can write $\beta\approx- 2 \pi \xi \alpha \frac{\ex{\edb}}{3 H \dot{\chi}}$ and the estimate \eqref{zeta3est} directly gives the generalization of \eqref{psback}
\be
\ex{\zeta(\vec{x})^3} \simeq \frac{1}{4\pi^3 \xi^3}. \label{zeta3st}
\ee

Finally we want to convert these results into a value for $f_{\rm NL}$. We take $f_{\rm NL}$ to be defined via
\ba
\ex{\zeta(\vec k_{1})\zeta(\vec k_{2})\zeta(\vec k_{3})}&=& (2\pi)^3 \delta^{3} \left(\vec{k}_{1}+\vec{k}_{2}+\vec{k}_{3}\right)\nn\\
&& \qquad \times (2\pi)^4 \frac{3}{10} f_{\rm NL}\Delta_{\zeta}^{4}(k) \frac{\sum_i k_i^3}{\Pi_i k_i^3}.\nn\\
&&
\ea
This gives
\be
f_{\rm NL} = B(\vec k_{1},\vec k_{2},\vec k_{3}) \frac{10}{3} \frac{1}{(2\pi)^4} \frac{1}{\Delta_{\zeta}^{4}(k)} \frac{\Pi_i k_i^3}{\sum_i k_i^3},
\ee
which for the equilateral case becomes
\ba
f_{\rm NL}^{\rm eq}& =& B_{\rm eq}(\vec{k}) \frac{10}{3} \frac{1}{(2\pi)^4} \frac{1}{\Delta_{\zeta}^{4}(k)} \frac{k^9}{3 k^3}\nn\\
&=& \frac{(2\pi)^6}{8\pi^2} \frac{1}{k^6}\ex{\zeta(\vec{x})^3} \times \frac{10}{3} \frac{1}{(2\pi)^4} \frac{1}{\Delta_{\zeta}^{4}(k)} \frac{k^9}{3 k^3}\nn\\
&=& \frac{10}{9} \frac{(2\pi)^2}{8\pi^2} \frac{\ex{\zeta(\vec{x})^3}}{\Delta_{\zeta}^{4}(k)}. \label{fnldef}
\ea
In the regime of negligible backreaction we can then take our estimate \eqref{well}, and conclude that
\be
f^{\rm eq}_{\rm NL}=\frac{2.8\cdot 10^{-7}}{\xi^9}  \frac{e^{6\pi\xi} \Delta_{\zeta,{\rm sr}}^{6}(k)}{\Delta_{\zeta}^{4}(k)}.
\ee
This again matches the result obtained in \cite{Barnaby:2010vf,Barnaby:2011vw} by a more precise computation. (Of course, after that we had found that the expressions for $\ex{\zeta(\vec{x})^3}$ match so well, this is only a consistency check.)

In the regime of strong backreaction, finally, we need to insert \eqref{zeta3st} into \eqref{fnldef}. Using our power spectrum estimate \eqref{psback} we find
\ba
f_{\rm NL}^{\rm eq} & =& \frac{10}{9} \frac{(2\pi)^2}{8\pi^2}  \frac{(2\pi\xi)^4}{4\pi^3\xi^3} = \frac{10}{9} 2\pi \xi \simeq 42 \label{fnlstr}
\ea
where we have used that towards the end of inflation we have $\xi\simeq6$.

Notwithstanding the precise match between \eqref{well} and \eqref{z3est}, there is a still an order one factor between the estimate for the three point function (and for $f_{\rm NL}$) and its precisely computed numerical value. Again: to arrive at \eqref{well} we have used the estimate \ref{edbbar} for $\ex{\edb}$, and to arrive at \eqref{z3est} we have inserted the large $\xi$ approximation $\frac{2.8\cdot 10^{-7}}{\xi^9}$ for $f_3(\xi)$. When using precise numerical prescriptions rather than estimates for $\ex{\edb}$ and $f_3(\xi)$ we find that our estimates overshoots the precisely computed $f_{\rm NL}$ by a factor of 9.5 at $\xi=2.2$ ($N=60$), and by a factor of 3.8 at $\xi=3$ ($N\approx 35$).

\begin{figure}[t!]
\begin{center}
\includegraphics[scale=0.35]{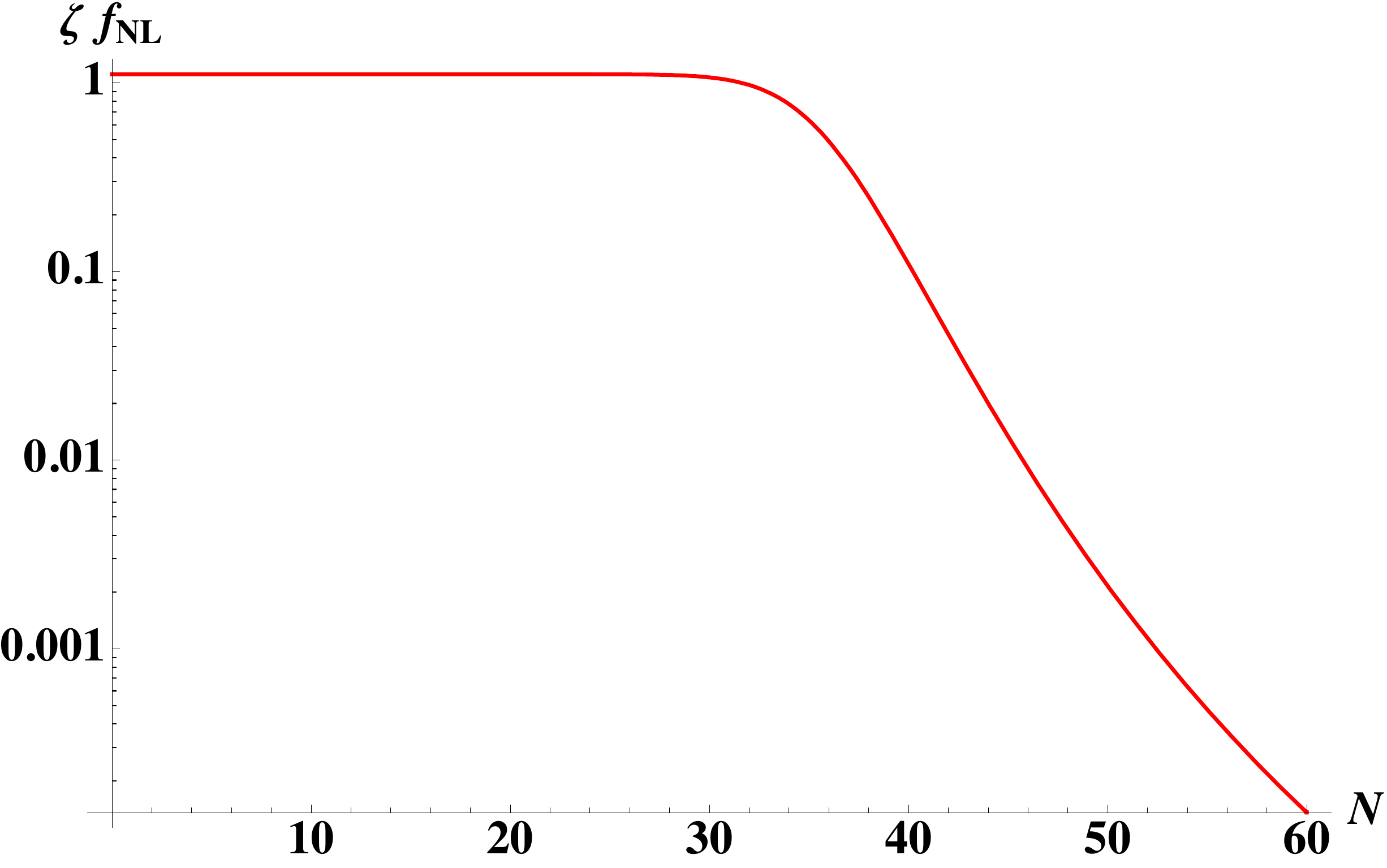}
\end{center}
\caption{Evolution of $f_{\rm NL}\times \zeta$ as a function of $N$, for $\xi[N=60]=2.2$}
\label{fnlzetaplot}
\end{figure}

Again we will not bother introducing a fudge factor to close this gap at some preferred value of $\xi$. Anyway, when backreaction is large $f_{\rm NL}$ is not a suitable indicator for the amount of non-Gaussianity anymore. In Figure \ref{fnlzetaplot} we plot our estimate for a more meaningful quantity: the skewness, which is equivalent to $f_{\rm NL}\zeta$. When backreaction becomes important, it saturates at a value of about one, which a posteriori justifies our approach \eqref{ngapp}.

\section{Black hole masses}\label{a:bh}

In this appendix we give some details about the derivation of \eqref{enrbh} for the black hole mass and about the total number of efoldings enforced by a specific expansion history.

Suppose the universe is radiation dominated right after the end of inflation. Then the expansion proceeds as $a \sim ({t/t_{0}})^{{1/2}}$, so $H(t) = {1\over 2 t}$. This regime starts at $t_{0}$, which is not the time since the beginning of Big Bang, but simply the constant $t_{0} = {1\over 2H}$, where $H$ is the Hubble constant at the end of inflation. We distinguish it from the decreasing $H(t)  = {1\over 2 t}$. The wavelength $l_{t_{0}}=H^{-1}e^{N}$ grows as $l_{t}=H^{-1} (t/t_{0})^{1/2} e^{N} = H^{-1} (2Ht)^{1/2} e^{N}$. The horizon size $1/H(t) = 2t$ grows and becomes equal to $l_{t}$ (and black holes form) at 
$$
2t = H^{-1}   (2Ht)^{1/2} e^{N}
$$
i.e. at 
$$
(2Ht)^{1/2} =(t/t_{0})^{1/2} = e^{N}.
$$
In other words, the black holes form after the universe expands by a factor $e^N$ since the end of inflation. The initial energy stored inside the volume $H^{-1}e^{N}$ was $M_{N}\simeq 10 \,e^{3N}$ g, but during this extra expansion it scales down (redshifts) by the factor $e^{N}$, so it becomes
$$
M_{BH}\simeq 10 \,e^{2N} g\,.
$$
It should be stressed that specifying the energy density at the end of inflation and at reheating, directly determines the number of efoldings corresponding to any scale (and in particular CMB scales) according to \cite{LL}
\be
N(k)&=&62-\log \frac{k}{a_{0}H_{0}}-\log \frac{10^{16}{\rm GeV}}{V_{\ast}^{1/4}}\\
&&\quad +\log \frac{V_{\ast}^{1/4}}{V_{end}^{1/4}}-\frac{1}{3}\log\frac{V_{end}^{1/4}}{\rho_{reh}^{1/4}}\,,\nonumber
\ee
where $V_{\ast}$ is the energy density during inflation when the mode $k$ left the horizon, $V_{end}$ is the energy density at the end of inflation, $\rho_{reh}$ is the energy density at reheating and the subscript $0$ refers to today's value. Taking for example $\rho_{reh}=V_{end}=m^{2}\Mpl^{2}/2$ and $V_{k}=m^{2}15^{2}\Mpl^{2}/2$ with $m=6\times 10^{6} \Mpl$ gives $N_{CMB}=N(a_{0}H_{0})\simeq 64$. We use this value in our discussion of primordial black holes, but since the difference between $60$ and $64$ changes very little in our numerics, for simplicity we use $N_{CMB}=60$ in the rest of the paper.


\begin{thebibliography}{99}


\bibitem{Kallosh:2010ug} 
  R.~Kallosh and A.~Linde,
  ``New models of chaotic inflation in supergravity,''
  JCAP {\bf 1011}, 011 (2010)
  [arXiv:1008.3375 [hep-th]].
  
\bibitem{Kallosh:2010xz} 
  R.~Kallosh, A.~Linde and T.~Rube,
  ``General inflaton potentials in supergravity,''
  Phys.\ Rev.\ D {\bf 83}, 043507 (2011)
  [arXiv:1011.5945 [hep-th]].
  
 
\bibitem{Kallosh:2011qk} 
  R.~Kallosh, A.~Linde, K.~A.~Olive and T.~Rube,
``Chaotic inflation and supersymmetry breaking,''
  Phys.\ Rev.\ D {\bf 84}, 083519 (2011)
  [arXiv:1106.6025 [hep-th]].
  

\bibitem{Kawasaki:2000yn}
  M.~Kawasaki, M.~Yamaguchi and T.~Yanagida,
  ``Natural chaotic inflation in supergravity,''
  Phys.\ Rev.\ Lett.\  {\bf 85}, 3572 (2000)
  [arXiv:hep-ph/0004243].
   

\bibitem{Yamaguchi:2001pw}
  M.~Yamaguchi,
  ``Natural double inflation in supergravity,''
  Phys.\ Rev.\  D {\bf 64}, 063502 (2001)
  [arXiv:hep-ph/0103045].
  
  
\bibitem{Kawasaki:2001as}
  M.~Kawasaki and M.~Yamaguchi,
``A supersymmetric topological inflation model,''
  Phys.\ Rev.\  D {\bf 65}, 103518 (2002)
  [arXiv:hep-ph/0112093].

\bibitem{Yamaguchi:2003fp}
  M.~Yamaguchi and J.~Yokoyama,
``Chaotic hybrid new inflation in supergravity with a running spectral
 index,''
  Phys.\ Rev.\  D {\bf 68} (2003) 123520
  [arXiv:hep-ph/0307373].
  
\bibitem{Brax:2005jv}
  P.~Brax and J.~Martin,
``Shift symmetry and inflation in supergravity,''
  Phys.\ Rev.\  D {\bf 72}, 023518 (2005)
  [arXiv:hep-th/0504168].
  
\bibitem{Kallosh:2007ig}
  R.~Kallosh,
  ``On inflation in string theory,''
  arXiv:hep-th/0702059.
  


\bibitem{Kadota:2007nc}
  K.~Kadota and M.~Yamaguchi,
``D-term chaotic inflation in supergravity,''
  Phys.\ Rev.\  D {\bf 76}, 103522 (2007)
  [arXiv:0706.2676 [hep-ph]].
  
\bibitem{Davis:2008fv}
  S.~C.~Davis and M.~Postma,
``SUGRA chaotic inflation and moduli stabilisation,''
  JCAP {\bf 0803}, 015 (2008)
  [arXiv:0801.4696 [hep-ph]].
  
\bibitem{Einhorn:2009bh}
  M.~B.~Einhorn and D.~R.~T.~Jones,
``Inflation with Non-minimal Gravitational Couplings in Supergravity,''
  JHEP {\bf 1003}, 026 (2010)
  [arXiv:0912.2718 [hep-ph]].


\bibitem{Ferrara:2010yw}
  S.~Ferrara, R.~Kallosh, A.~Linde, A.~Marrani and A.~Van Proeyen,
``Jordan Frame Supergravity and Inflation in NMSSM,'' Phys. Rev. {\bf D82}, 045003
(2010)
  [arXiv:1004.0712 [hep-th]].

\bibitem{Lee:2010hj}
  H.~M.~Lee,
``Chaotic inflation in Jordan frame supergravity,''
  JCAP {\bf 1008}, 003 (2010)
  [arXiv:1005.2735 [hep-ph]].

\bibitem{Ferrara:2010in}
  S.~Ferrara, R.~Kallosh, A.~Linde, A.~Marrani and A.~Van Proeyen,
``Superconformal Symmetry, NMSSM, and Inflation,''
  arXiv:1008.2942 [hep-th].


\bibitem{Takahashi:2010ky}
  F.~Takahashi,
``Linear Inflation from Running Kinetic Term in Supergravity,''
  arXiv:1006.2801 [hep-ph];
  
\bibitem{Nakayama:2010kt}
  K.~Nakayama and F.~Takahashi,
``Running Kinetic Inflation,''
  arXiv:1008.2956 [hep-ph].

  
\bibitem{Silverstein:2008sg}
  E.~Silverstein and A.~Westphal,
 ``Monodromy in the CMB: Gravity Waves and String Inflation,''
  Phys.\ Rev.\  D {\bf 78}, 106003 (2008)
  [arXiv:0803.3085 [hep-th]].
  
\bibitem{McAllister:2008hb}
  L.~McAllister, E.~Silverstein and A.~Westphal,
``Gravity Waves and Linear Inflation from Axion Monodromy,''
  arXiv:0808.0706 [hep-th].




\bibitem{Yamaguchi:2000vm}
  M.~Yamaguchi and J.~Yokoyama,
``New inflation in supergravity with a chaotic initial condition,''
  Phys.\ Rev.\  D {\bf 63}, 043506 (2001)
  [arXiv:hep-ph/0007021].
  
\bibitem{Kallosh:2004yh}
  R.~Kallosh and A.~D.~Linde,
``Landscape, the scale of SUSY breaking, and inflation,''
  JHEP {\bf 0412}, 004 (2004)
  [arXiv:hep-th/0411011];
  J.~J.~Blanco-Pillado, R.~Kallosh and A.~D.~Linde,
``Supersymmetry and stability of flux vacua,''
  JHEP {\bf 0605}, 053 (2006)
  [hep-th/0511042];
  R.~Kallosh and A.~D.~Linde,
``Testing String Theory with CMB,''
  JCAP {\bf 0704}, 017 (2007)
  [arXiv:0704.0647 [hep-th]].

\bibitem{Linde:2011ja} 
  A.~Linde, Y.~Mambrini and K.~A.~Olive,
``Supersymmetry Breaking due to Moduli Stabilization in String Theory,''
  Phys.\ Rev.\ D {\bf 85}, 066005 (2012)
  [arXiv:1111.1465 [hep-th]].
  
\bibitem{Dudas:2012wi} 
  E.~Dudas, A.~Linde, Y.~Mambrini, A.~Mustafayev and K.~A.~Olive,
``Strong moduli stabilization and phenomenology,''
  arXiv:1209.0499 [hep-ph].
  
  
  
\bibitem{Ferrara:2011dz} 
  S.~Ferrara and R.~Kallosh,
``Creation of Matter in the Universe and Groups of Type E7,''
  JHEP {\bf 1112}, 096 (2011)
  [arXiv:1110.4048 [hep-th]].


\bibitem{curva}  A.~D.~Linde and V.~Mukhanov,  ``Nongaussian isocurvature
perturbations from inflation,''  Phys.\ Rev.\ D {\bf 56}, 535 (1997)
[arXiv:astro-ph/9610219];  K.~Enqvist and M.~S.~Sloth,
``Adiabatic CMB perturbations in pre big bang string cosmology,''
  Nucl.\ Phys.\  B {\bf 626}, 395 (2002)
  [arXiv:hep-ph/0109214]; D.~H.~Lyth and D.~Wands, ``Generating the curvature
perturbation without an inflaton,''  Phys.\ Lett.\ B {\bf 524}, 5 (2002)
[arXiv:hep-ph/0110002];  T.~Moroi and T.~Takahashi, ``Effects of cosmological
moduli fields on cosmic microwave background,''  Phys.\ Lett.\ B {\bf 522}, 215 (2001)  [Erratum-ibid.\ B {\bf 539}, 303 (2002)]
[arXiv:hep-ph/0110096].

  
\bibitem{Demozzi:2010aj} 
  V.~Demozzi, A.~Linde and V.~Mukhanov,
``Supercurvaton,''
  JCAP {\bf 1104}, 013 (2011)
  [arXiv:1012.0549 [hep-th]].



\bibitem{Anber:2006xt} 
  M.~M.~Anber and L.~Sorbo,
``N-flationary magnetic fields,''
  JCAP {\bf 0610}, 018 (2006)
  [astro-ph/0606534].
  
\bibitem{Anber:2009ua} 
  M.~M.~Anber and L.~Sorbo,
 ``Naturally inflating on steep potentials through electromagnetic dissipation,''
  Phys.\ Rev.\ D {\bf 81}, 043534 (2010)
  [arXiv:0908.4089 [hep-th]].
  
\bibitem{Durrer:2010mq} 
  R.~Durrer, L.~Hollenstein and R.~K.~Jain,
``Can slow roll inflation induce relevant helical magnetic fields?,''
  JCAP {\bf 1103}, 037 (2011)
  [arXiv:1005.5322 [astro-ph.CO]].
  
\bibitem{Barnaby:2010vf}
N.~Barnaby and M.~Peloso,
``Large Nongaussianity in Axion Inflation,''
Phys.\ Rev.\ Lett.\ {\bf 106} (2011) 181301
[arXiv:1011.1500 [hep-ph]].

  
\bibitem{Barnaby:2011vw} 
  N.~Barnaby, R.~Namba and M.~Peloso,
``Phenomenology of a Pseudo-Scalar Inflaton: Naturally Large Nongaussianity,''
  JCAP {\bf 1104}, 009 (2011)
  [arXiv:1102.4333 [astro-ph.CO]].
  


  
\bibitem{Barnaby:2011qe} 
  N.~Barnaby, E.~Pajer and M.~Peloso,
 ``Gauge Field Production in Axion Inflation: Consequences for Monodromy, non-Gaussianity in the CMB, and Gravitational Waves at Interferometers,''
  Phys.\ Rev.\ D {\bf 85}, 023525 (2012)
  [arXiv:1110.3327 [astro-ph.CO]].


\bibitem{Meerburg:2012id} 
  P.~D.~Meerburg and E.~Pajer,
``Observational Constraints on Gauge Field Production in Axion Inflation,''
  arXiv:1203.6076 [astro-ph.CO].
  
  
\bibitem{SC}
J.~L.~Cook and L.~Sorbo,
``Particle Production during Inflation and Gravitational Waves Detectable by Ground-Based Interferometers,''
Phys.\ Rev.\ D {\bf 85} (2012) 023534
[Erratum-ibid.\ D {\bf 86} (2012) 069901]
[arXiv:1109.0022 [astro-ph.CO]].


\bibitem{AS}
M.~M.~Anber and L.~Sorbo,
``Non-Gaussianities and Chiral Gravitational Waves in Natural Steep Inflation,''
Phys.\ Rev.\ D {\bf 85} (2012) 123537
[arXiv:1203.5849 [astro-ph.CO]].
N.~Barnaby, J.~Moxon, R.~Namba, M.~Peloso, G.~Shiu and P.~Zhou,
``Gravity Waves and Non-Gaussian Features from Particle Production in a Sector Gravitationally Coupled to the Inflaton,''
Phys.\ Rev.\ D {\bf 86} (2012) 103508
[arXiv:1206.6117 [astro-ph.CO]].


\bibitem{malik}
  A.~S.~Josan, A.~M.~Green and K.~A.~Malik,
  ``Generalised constraints on the curvature perturbation from primordial black holes,''
  Phys.\ Rev.\ D {\bf 79} (2009) 103520
  [arXiv:0903.3184 [astro-ph.CO]].
  
\bibitem{Carr:2009jm} 
  B.~J.~Carr, K.~Kohri, Y.~Sendouda and J.~'i.~Yokoyama,
  ``New cosmological constraints on primordial black holes,''
  Phys.\ Rev.\ D {\bf 81}, 104019 (2010)
  [arXiv:0912.5297 [astro-ph.CO]].
  
\bibitem{Byrnes:2012yx} 
  C.~T.~Byrnes, E.~J.~Copeland and A.~M.~Green,
  ``Primordial black holes as a tool for constraining non-Gaussianity,''
  Phys.\ Rev.\ D {\bf 86}, 043512 (2012)
  [arXiv:1206.4188 [astro-ph.CO]].
  
\bibitem{Klimai:2012sf} 
E.~Bugaev and P.~Klimai,
``Curvature perturbation spectra from waterfall transition, black hole constraints and non-Gaussianity,''
  JCAP {\bf 1111}, 028 (2011)
  [arXiv:1107.3754 [astro-ph.CO]].
E.~Bugaev and P.~Klimai,
``Formation of primordial black holes from non-Gaussian perturbations produced in a waterfall transition,''
  Phys.\ Rev.\ D {\bf 85}, 103504 (2012)
  [arXiv:1112.5601 [astro-ph.CO]].
 P.~A.~Klimai and E.~V.~Bugaev,
``Primordial black hole formation from non-Gaussian curvature perturbations,''
  arXiv:1210.3262 [astro-ph.CO].
  
\bibitem{Lyth:2012yp} 
  D.~H.~Lyth,
  ``The hybrid inflation waterfall and the primordial curvature perturbation,''
  JCAP {\bf 1205}, 022 (2012)
  [arXiv:1201.4312 [astro-ph.CO]].
  
  
  
\bibitem{Shandera:2012ke} 
  S.~Shandera, A.~L.~Erickcek, P.~Scott and J.~Y.~Galarza,
  ``Number Counts and Non-Gaussianity,''
  arXiv:1211.7361 [astro-ph.CO].
  
  



  
    \bibitem{gravitino}
  P.~Fayet, ``Phenomenology Of Supersymmetry,''
  Talk at the XVIIth Rencontre de Moriond, Ecole Normale Superieure preprint LPTENS 82/10 (1982);
S.~Weinberg,
``Cosmological Constraints on the Scale of Supersymmetry Breaking,''
  Phys.\ Rev.\ Lett.\  {\bf 48}, 1303 (1982);
J.~R.~Ellis, A.~D.~Linde and D.~V.~Nanopoulos,
``Inflation Can Save the Gravitino,''
  Phys.\ Lett.\  B {\bf 118}, 59 (1982); L.~M.~Krauss,
``New Constraints on Ino Masses from Cosmology. 1. Supersymmetric Inos,''
  Nucl.\ Phys.\  {\bf B227}, 556 (1983);
J.~R.~Ellis, J.~S.~Hagelin, D.~V.~Nanopoulos, K.~A.~Olive and M.~Srednicki,
 ``Supersymmetric Relics from the Big Bang,''
  Nucl.\ Phys.\  B {\bf 238}, 453 (1984);
 M.~Y.~Khlopov, A.~D.~Linde,
``Is It Easy to Save the Gravitino?,''
  Phys.\ Lett.\  {\bf B138}, 265-268 (1984).
  
  

  
\bibitem{Komatsu:2010fb} 
  E.~Komatsu {\it et al.}  [WMAP Collaboration],
``Seven-Year Wilkinson Microwave Anisotropy Probe (WMAP) Observations: Cosmological Interpretation,''
  Astrophys.\ J.\ Suppl.\  {\bf 192}, 18 (2011)
  [arXiv:1001.4538 [astro-ph.CO]].
  
\bibitem{Banks:2003sx}
T.~Banks, M.~Dine, P.~J.~Fox and E.~Gorbatov,
``On the Possibility of Large Axion Decay Constants,''
JCAP {\bf 0306} (2003) 001
[hep-th/0303252].




   
\bibitem{ng}
  C.~-M.~Lin and K.~-W.~Ng,
  ``Primordial Black Holes from Passive Density Fluctuations,''
  arXiv:1206.1685 [hep-ph].
  
  
\bibitem{Dimopoulos:2012av} 
  K.~Dimopoulos and M.~Karciauskas,
``Parity Violating Statistical Anisotropy,''
  JHEP {\bf 1206}, 040 (2012)
  [arXiv:1203.0230 [hep-ph]].
 
 
   \bibitem{Linde:1983gd} A.~D.~Linde, ``Chaotic Inflation,'' Phys.\ Lett.\ B 
{\bf 129}, 177 (1983). 






\bibitem{split}
 N.~Arkani-Hamed and S.~Dimopoulos,
``Supersymmetric unification without low energy supersymmetry and signatures
for fine-tuning at the LHC,''
  JHEP {\bf 0506}, 073 (2005)
  [arXiv:hep-th/0405159];
  G.~F.~Giudice and A.~Strumia,
``Probing High-Scale and Split Supersymmetry with Higgs Mass Measurements,''
  Nucl.\ Phys.\ B {\bf 858}, 63 (2012)
  [arXiv:1108.6077 [hep-ph]];
   M.~Ibe, S.~Matsumoto and T.~T.~Yanagida,
``Pure Gravity Mediation with $m_{3/2}$ = 10-100TeV,''
  Phys.\ Rev.\ D {\bf 85}, 095011 (2012)
  [arXiv:1202.2253 [hep-ph]].
  
\bibitem{LL}
A.~R.~Liddle and D.~H.~Lyth,
``The Cold Dark Matter Density Perturbation,''
Phys.\ Rept.\ {\bf 231} (1993) 1
[astro-ph/9303019].

\end{thebibliography}
\end{document}